\documentclass[prb,preprint]{revtex4} 

\usepackage{amsmath}  
\usepackage{amsfonts} 
\usepackage{graphicx} 

\graphicspath{%
    {converted_graphics/}
    {/}
}
\begin{document}


\title{The Delicate Dance of Orbital Rendezvous}

\author{Bradley W.\ Carroll}
\email{bcarroll@weber.edu} 
\affiliation{Department of Physics, Weber State University, Ogden, UT  84408-2508}



\date{\today}

\begin{abstract}
The meeting of two spacecraft in orbit around a planet or moon involves a delicate dance that must carefully the balance the gravitational, Coriolis, and centrifugal forces acting on the spacecraft.  The intricacy of the relative motion between the two spacecraft caused problems for the Gemini missions in the mid-1960s.  Although now mastered, the problem of how to bring two orbiting objects together continues to be misrepresented in popular movies and books.  In this article, I will consider the case when the two spacecraft are in close proximity (compared with the radii of their orbits), and examine the counter-intuitive trajectories that are needed to bring them together.  I will examine how a stranded astronaut might use an impulsive force to return to her ship in Earth orbit, how and when line-of-sight targeting may be used for a rendezvous, and how the Apollo 11 lunar module executed a Terminal Phase Initiation maneuver to rendezvous with the command/service module as they both circled the Moon.
\end{abstract}

\maketitle 

\section{Introduction} 

Fifty years ago, Neil Armstrong and Buzz Aldrin landed their lunar module onto the Moon while Michael Collins orbited above in the command/service module.  The upper stage of the lunar module lifted off from the Moon 21.6~h later and the two spacecraft rendezvoused to reunite their crews.  NASA historian James R.\ Hansen has written, ``Thousands of factors contributed to the ultimate success of Apollo, but no single factor was more essential than the concept of lunar-orbit rendezvous.''\cite{Hansen}  An orbital rendezvous is not simple, even when the two spacecraft are relatively close together.  During the 1965 flight of Gemini IV and its attempted rendezvous with a booster target, astronaut James ``McDivitt braked the spacecraft, aimed it, and thrusted toward the target. After two bursts from his thrusters, the booster seemed to move away and downward. A few minutes later, McDivitt pitched the spacecraft nose down and the crew again saw the rocket, which seemed to be traveling on a different track. He thrusted toward it -- no success --
and stopped. McDivitt repeated this sequence several times with the same luck.''\cite{HackerGrimwood1}  Later that year, after Gemini VI-A's successful rendezvous with Gemini VII, astronaut Wally Schirra stated, ``Somebody said\ldots when you come to within three miles [five kilometers], you've rendezvoused. If anybody thinks they've pulled a rendezvous off at three miles, have fun! This is when we started doing our work. I don't think rendezvous is over until you are stopped -- completely stopped -- with no relative motion between the two vehicles, at a range of approximately 120 feet [about 40 meters].  That's rendezvous! From there on, it's stationkeeping. That's when you can go back and play the game of driving a car or driving an airplane or pushing a skateboard -- it's about that simple.''\cite{HackerGrimwood2}  The orbital mechanics during this interval, between being relatively close and being just a few meters away, is the subject of this paper.

From our perspective, 50 years later, the orbital mechanics of rendezvous are well understood.  However, the topic of orbital mechanics is covered only in specialized courses not found in most physics curricula.  The Hill and Clohessy-Wiltshire equations (derived below) may be found in texts,\cite{Wiesel,Prussing} but interesting properties of their solutions are not made explicit.  In this journal, the topic of relative orbital motion has been previously discussed in the context of the motion of an astronaut floating near a space station,\cite{Freedman} an object ejected from an orbiting space station,\cite{Butikov} and the orbital dynamics of the three LISA spacecraft.\cite{Amato}  In this paper I will describe and elaborate upon the results of a little-known 1962 Air Force document, written by Donald Mueller and now unclassified, on ``Relative Motion in the Docking Phase of Orbital Rendezvous.''\cite{Mueller}  In Sec.~\ref{sec:mechanics}, the Hill and Clohessy-Wiltshire equations will be derived, and the nature of the trajectory of an interceptor spacecraft near a target spacecraft will be described.  Section~\ref{sec:astronaut} deals with the problem of a stranded astronaut who must fire a thruster to return to her spacecraft.  In Sec.~\ref{sec:los} I investigate Shirra's claim that an orbital rendezvous is achieved when the interceptor and target are about 40~m apart, and find the distance of closest approach using line-of-sight targeting.  The paper closes with Sec.~\ref{sec:Apollo11}, which describes the historic docking of Apollo~11's lunar and command/service modules.  

\section{The Mechanics of Rendezvous}\label{sec:mechanics}
    
Consider a spacecraft in a circular orbit around Earth or the Moon, and call this spacecraft the {\em target}.  Another spacecraft, the {\em interceptor}, ascends to orbit near the target spacecraft.  How can the interceptor briefly fire its thrusters so it then coasts to a rendezvous with the target?  The obvious answer is to point the nose of the interceptor toward the target and create a short thrust toward the target, and like most obvious answers in orbital mechanics, it is wrong.  This can be clearly seen by imagining two spacecraft in the same circular orbit (same orbital plane and altitude), with the target leading the interceptor.  If the interceptor fires a forward thrust to increase its speed, it will rise to a higher orbit.  In this higher orbit, its angular velocity will actually decrease according to Kepler's third law, and it will fall farther behind the target.  One answer is for the interceptor to fire its thrusters to slow down, thus dropping into a lower orbit.  From there, its increased angular velocity will allow it to catch up with the target.  (The speeding-up effect also occurs with atmospheric drag on a satellite.\cite{Mills})  A carefully timed forward thrust will then increase its speed and raise the interceptor to meet the target.  An alternative requiring a single thrust is described in Sec.~\ref{sec:los}.

\begin{figure}[h] 
  \centering
  \includegraphics[width=3.701in,height=3.00in,keepaspectratio]{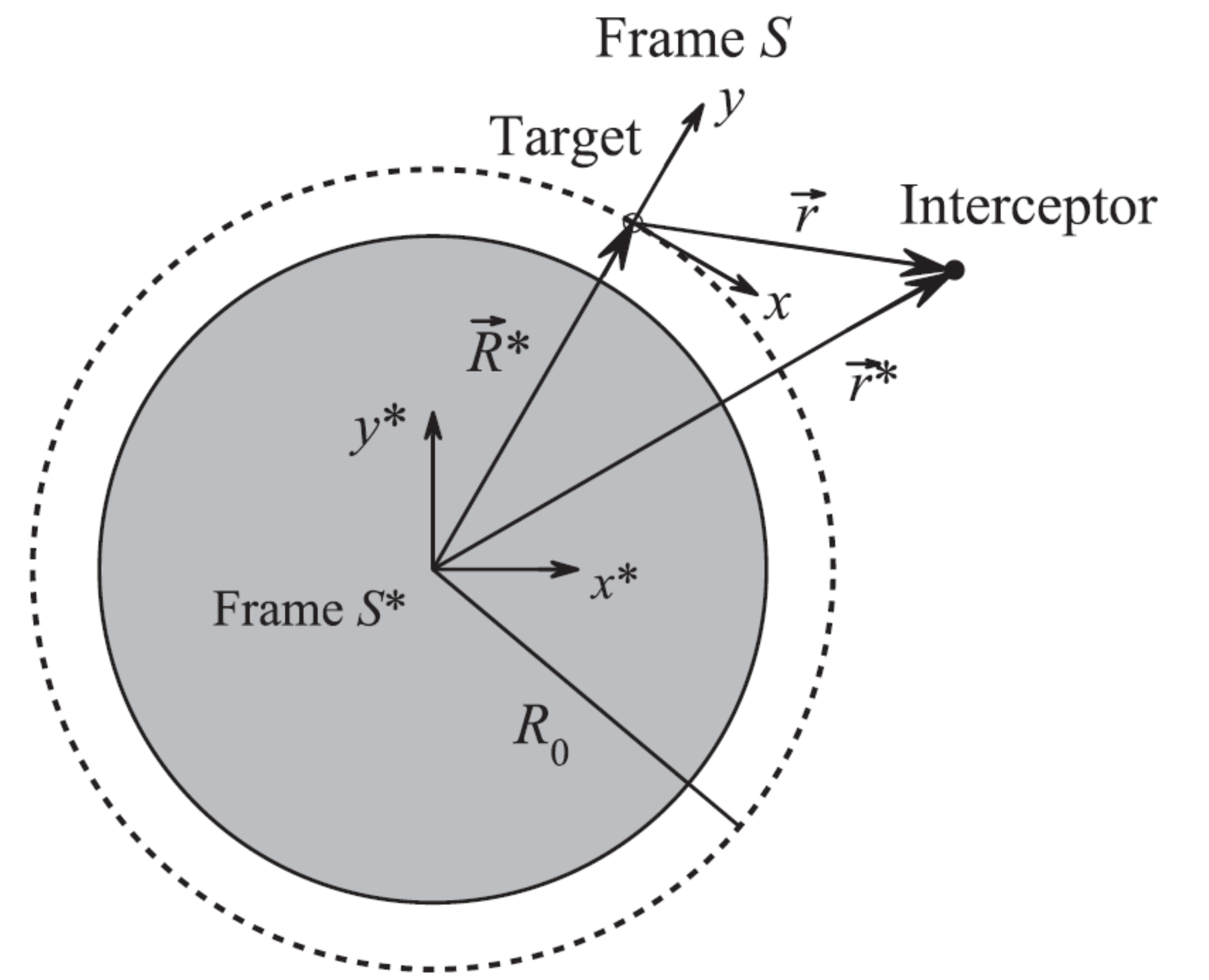} 
  \caption{Inertial reference frame $S^*$ is centered on Earth or the Moon, and the non-inertial reference frame $S$ is centered on the target spacecraft as it travels around the the massive central body with constant angular velocity $\omega_0$.  The $z^*$ and $z$ axes are directed out of the page; they are not shown for clarity.}
\label{fig:axes}
\end{figure}

To approach the problem of rendezvous systematically, let us adopt an inertial reference frame, fixed with respect to the stars and called $S^*$, with its origin at the center of the massive central body (for now, Earth; later, the Moon); see Fig.\ref{fig:axes}.  (The orbital motion of Earth around the Sun, and the Moon around Earth, may be safely ignored.)  The circular orbit of the target lies in the $x^*$-$y^*$ plane of $S^*$.  A non-inertial reference frame, $S$, has its origin at the center-of-mass of the target spacecraft with the $x$-axis in the direction of the target's orbital motion.  As the target spacecraft travels around Earth, it also rotates to maintain its orientation relative to the ground.  Thus the frame $S$ rotates about the $z$-axis so that the $y$-axis always points radially away from the origin of $S^*$.  The location of the target is
\begin{equation}
\vec{R^*}=X^*\,\hat{i^*}+Y^*\,\hat{j^*}=R_0\hat{j},
\label{eq:Rstar}
\end{equation}
where $R_0=\sqrt{X^{*2}+Y^{*2}}$ is the radius of the target's orbit.  As seen from inertial frame $S^*$, the location of the interceptor is 
\begin{equation}
\vec{r^*}=x^*\,\hat{i^*}+y^*\,\hat{j^*}+z^*\,\hat{k^*},
\label{eq:rstar}
\end{equation}
and as seen from the target frame $S$, the location of the interceptor is 
\begin{equation}
\vec{r}=x\,\hat{i}+y\,\hat{j}+z\,\hat{k}.
\label{eq:r}
\end{equation}
The goal is to describe the motion of the interceptor as seen from the target frame $S$ when the two spacecraft are close, i.e., when $x$, $y$, and $z\ll R_0$.  (The nonlinear terms that arise when the two spacecraft are not close are neglected.\cite{Omran,Lovell,Stringer})   

The relations between the position, velocity, and acceleration of the interceptor measured from the frames $S^*$ and $S$ are given by\cite{Symon}
\begin{eqnarray}
\vec{r}&=&\vec{r^*}-\vec{R^*}\label{eq:r}\\
\dfrac{d\vec{r}}{dt}&=&\dfrac{d^*\vec{r^*}}{dt}-\vec{\omega}\times \vec{r}-\dfrac{d^*\vec{R^*}}{dt}\label{eq:v}\\
\dfrac{d^2\vec{r}}{dt^2}&=&\dfrac{d^{*2}\vec{r^*}}{dt^2}-\vec{\omega}\times (\vec{\omega}\times\vec{r})-2\vec{\omega}\times\dfrac{d\vec{r}}{dt}-\dfrac{d^*\vec{\omega}}{dt}\times\vec{r}-\dfrac{d^{*2}\vec{R^*}}{dt^2},\label{eq:a}
\end{eqnarray}
where derivatives with respect to frame $S^*$ are indicated with a star following the ``d'' in the numerator, and derivatives with respect to frame $S$ have no such star.  Newton's second law of motion applied to the interceptor in the inertial frame $S^*$ is
\begin{equation}
m_i\dfrac{d^{*2}\vec{r^*}}{dt^2}=-\dfrac{GMm_i}{r^{*3}}\vec{r^*}+\vec{F},
\label{eq:Newton2}
\end{equation}   
where $m_i$ is the mass of the interceptor and $\vec{F}$ is the sum of other (non-gravitational) forces acting on the interceptor.  (These non-gravitational forces will be set equal to zero below, but are included here for completeness.)  The magnitude of the angular velocity vector, $\vec{\omega}_0$, comes from Kepler's third law relating the orbital period $P$ and the semimajor axis $a^*$ of the target's orbit,
\begin{equation}
P^2=\dfrac{4\pi^2a^{*3}}{G(M+m_t)},
\label{eq:Kepler3rdgen}
\end{equation}
where $M$ is the mass of the central body and $m_t$ is the (negligibly small) mass of the target spacecraft.\cite{Carroll}  For the target in a circular orbit, $a^*=R_0$, so
\begin{equation}
P^2=\left(\dfrac{2\pi}{\omega_0}\right)^2=\dfrac{4\pi^2R_0^3}{GM},
\end{equation}
where $m_t$ has been ignored.  Thus 
\begin{equation}
\omega_0=\sqrt{\dfrac{GM}{R_0^3}}
\label{eq:omegamag}
\end{equation}
and the angular velocity vector of frame $S$ as seen from frame $S^*$ is given by
\begin{equation}
\vec{\omega}=-\omega_0\hat{k}.
\label{eq:omegavec}
\end{equation}
Because the angular velocity vector is constant, the fourth term on the right-hand side of Eq.~(\ref{eq:a}) is zero.  The other terms in this equation are
the non-inertial centrifugal force per unit mass on the interceptor in frame $S$,
\begin{equation}
-\vec{\omega}\times (\vec{\omega}\times \vec{r})=\omega_0^2x\hat{i}+\omega_0^2y\hat{j}=\omega_0^2\vec{r}-\omega_0^2z\hat{k},
\label{eq:centrifugal}
\end{equation}
and the centripetal acceleration of the target in the inertial frame $S^*$,
\begin{equation}
\dfrac{d^{*2}\vec{R^*}}{dt^2}=-\omega_0^2\vec{R^*}=-\dfrac{GM}{R_0^3}\vec{R^*}.
\label{eq:Saccel}
\end{equation}
Inserting these into Eq.~(\ref{eq:a}) and utilizing Eqs.~(\ref{eq:r}), (\ref{eq:Newton2}), and (\ref{eq:omegamag}) leads to
\begin{equation}
\dfrac{d^2\vec{r}}{dt^2}=\left(\dfrac{GM}{R_0^3}-\dfrac{GM}{r^{*3}}\right)\vec{r^*}-\omega_0^2z\hat{k}-2\vec{\omega}\times\dfrac{d\vec{r}}{dt}+\dfrac{\vec{F}}{m_i},
\label{eq:almost}
\end{equation}
or alternatively,
\begin{equation}
\dfrac{d^2\vec{r}}{dt^2}=\left(1-\dfrac{R_0^3}{r^{*3}}\right)\omega_0^2\vec{r^*}-\omega_0^2z\hat{k}-2\vec{\omega}\times\dfrac{d\vec{r}}{dt}+\dfrac{\vec{F}}{m_i}.
\label{eq:almost2}
\end{equation}
In order to linearize Eq.~(\ref{eq:almost2}), let us assume that the interceptor is close to the target.  The first term on the right-hand side is, using Eq.~(\ref{eq:r}),
\begin{eqnarray}
\left(1-\dfrac{R_0^3}{r^{*3}}\right)\omega_0^2\vec{r^*}&=&\left\{1-\dfrac{R_0^3}{[x^2+(R_0+y)^2+z^2]^{3/2}}\right\}\omega_0^2(\vec{r}+\vec{R^*})\nonumber\\
&=&\left\{1-\dfrac{1}{\left[\left(\dfrac{x}{R_0}\right)^2+\left(1+\dfrac{y}{R_0}\right)^2+\left(\dfrac{z}{R_0}\right)^2\right]^{3/2}}\right\}\omega_0^2(\vec{r}+R_0\hat{j}).\label{eq:term1almost}
\end{eqnarray}
Assuming that $x/R_0$, $y/R_0$, and $z/R_0\ll 1$, ignoring terms that are second-order in smallness, and expanding the fraction in a Taylor series results in
\begin{equation}
\left(1-\dfrac{R_0^3}{r^{*3}}\right)\omega_0^2\vec{r^*}\simeq 3\omega_0^2y\hat{j}\label{eq:term1}
\end{equation}
to first-order.  The third term on the right-hand side of Eq.~(\ref{eq:almost2}) is
\begin{equation}
-2\vec{\omega}\times\dfrac{d\vec{r}}{dt}=-2\omega_0\dfrac{dy}{dt}\hat{i}+2\omega_0\dfrac{dx}{dt}\hat{j}.
\label{eq:term2}
\end{equation}
Equation~(\ref{eq:almost2}) is then
\begin{equation}
\dfrac{d^2\vec{r}}{dt^2}=3\omega_0^2y\hat{j}-\omega_0^2z\hat{k}-2\omega_0\dfrac{dy}{dt}\hat{i}+2\omega_0\dfrac{dx}{dt}\hat{j}+\dfrac{\vec{F}}{m_i}.
\label{eq:CWvec}
\end{equation}
Looking at this equation and comparing it with Eq.~(\ref{eq:almost}), the first term on the right-hand side is a ``tidal'' term; it arises from the difference between the gravitational forces at the positions of the target and interceptor (at least to first order; $GM/R_0^3$ is multiplied by $\vec{r^*}$ instead of by $\vec{R^*}$).  The effect of the first term is to force the interceptor away from the $y=0$ plane.  Remaining at rest at $y=0$ is an unstable equilibrium.  The third and fourth terms in Eq.~(\ref{eq:CWvec}) -- and the third term in Eq.~(\ref{eq:almost}) -- are due to the Coriolis force, which acts in the $x$-$y$ plane perpendicular to the interceptor's velocity.

Let us express Eq.~(\ref{eq:CWvec}) in terms of its $x$, $y$, and $z$ components.  It will be convenient to change to dot notation for the time derivatives.  
\begin{eqnarray}
\ddot{x}&=&-2\omega_0\dot{y}+\dfrac{F_x}{m_i}\label{eq:CWx}\\
\ddot{y}&=&3\omega_0^2y+2\omega_0\dot{x}+\dfrac{F_y}{m_i}\label{eq:CWy}\\
\ddot{z}&=&-\omega_0^2z\label{eq:CWz}+\dfrac{F_z}{m_i}\label{eq:CWz}.
\end{eqnarray}
These are known as the Hill equations.\cite{Hill}  The forces $F_x$, $F_y$, and $F_z$ that appear on the right-hand sides will now be set equal to zero because it is assumed that only impulsive thrusts will be applied to the interceptor, which will coast afterward.  Note that the $z$ motion is not coupled to motion in the $x$ or $y$-directions, so it may be immediately solved (with $F_z=0$ and subject to the initial conditions that at $t=0$, $z=z_0$ and $\dot{z}=\dot{z}_0$) to find
\begin{equation}
z(t)=z_0\cos(\omega_0t)+\dfrac{\dot{z}_0}{\omega_0}\sin(\omega_0t)
\label{eq:zt}
\end{equation}
and
\begin{equation}
\dot{z}(t)=-\omega_0z_0\sin(\omega_0t)+\dot{z}_0\cos(\omega_0t).
\label{eq:zdott}
\end{equation}  
This describes simple harmonic motion along the $z$-axis of the target frame $S$.  It arises whenever the target and interceptor are orbiting in different planes.  These planes cross each other at two points at opposites ends of the orbits' common diameter.  The interceptor's thrusters must be fired in the $z$-direction until the two planes coincide, so the interceptor orbits in the $x$-$y$ plane of the target.  For the rest of this article, it is assumed that any such maneuvers have been successfully accomplished, so that $z(t)$ and $\dot{z}(t)$ have been reduced to zero.

The interplay of the Coriolis and ``tidal'' forces can be seen by examining a specific solution to Eqs.~(\ref{eq:CWx}) and (\ref{eq:CWy}), namely 
\begin{eqnarray}
x(t)&=&a\cos(\omega_0t)\label{eq:examplex}\\
y(t)&=&\frac{a}{2}\sin(\omega_0t),\label{eq:exampley}
\end{eqnarray}
as may be easily verified.  This describes the interceptor moving counterclockwise (as seen from a point on the positive $z$-axis) around an ellipse of semimajor axis $a$ in the $x$-$y$ plane.  Using Eq.~(\ref{eq:CWvec}) along with Eq.~(\ref{eq:omegavec}) for the angular velocity vector, the Coriolis force per unit mass acting on the interceptor is 
\begin{equation}
\vec{F}_{\rm Cor}=-2a\omega_0^2\left[\frac{1}{2}\cos(\omega_0t)\,\hat{i}+\sin{\omega_0t}\,\hat{j}\right]
\label{eq:FCoriolis}
\end{equation}
and the ``tidal'' force per unit mass is
\begin{equation}
\vec{F}_{\rm tidal}= \frac{3}{2}a\omega_0^2\sin(\omega_0t)\,\hat{j},
\label{eq:Ftidal}
\end{equation}
so the total force per unit mass acting on the interceptor is
\begin{equation}
\vec{F}_{\rm total}=\vec{F}_{\rm Cor}+\vec{F}_{\rm tidal}=-a\omega_0^2\left[ \cos(\omega_0t)\,\hat{i}+\frac{1}{2}\sin(\omega_0t)\,\hat{j}\right]=-\omega_0^2\vec{r},
\label{eq:Ftotal}
\end{equation}
where $\vec{r}$ is the position vector of the interceptor in frame $S$.  Thus, for this closed elliptical trajectory when the two spacecraft are close, the total force on the interceptor is always directed toward the origin and has a magnitude proportional to the distance from the target.  In other words, the total force in frame $S$ obeys Hooke's law.  It is not surprising that the Coriolis and ``tidal'' forces challenged the intuition of the early astronauts as they attempted to rendezvous with a target that acted as though it were attached to them by a spring.
 
Before finding a general solution to Eqs.~(\ref{eq:CWx}) and (\ref{eq:CWy}), let us find a conserved energy that can be employed in the non-inertial frame $S$.  The Coriolis force always acts perpendicular to the interceptor's motion, so it does no work and can be ignored.  The only work done on the interceptor is due to the ``tidal'' force in the $y$-direction.  This can be integrated to obtain a potential energy per unit mass,
\begin{equation}
U_m(y)=-\int 3\omega_0^2y\,dy=-\dfrac{3\omega_0^2y^2}{2}.
\label{eq:Um}
\end{equation}
Adding this to the kinetic energy per unit mass of the interceptor gives the conserved energy per unit mass,
\begin{equation}
E_m=\dfrac{1}{2}\left(\dot{x}^2+\dot{y}^2\right)-\dfrac{3\omega_0^2y^2}{2}=\mbox{constant}.
\label{eq:Em}
\end{equation}
The value of $\omega_0$ is about $10^{-3}$~rad/s for the International Space Station, so $y$ must be on the order of a kilometer to change the speed of the interceptor by as much as a meter per second.
 
Equations~(\ref{eq:CWx}) and (\ref{eq:CWy}) (with $F_x$ and $F_y=0$) can be solved subject to the usual initial conditions that at $t=0$, $x=x_0$ and $\dot{x}=\dot{x}_0$, and similarly for the $y$ initial conditions.
Integrating Eq.~(\ref{eq:CWx}) with respect to time and applying the boundary conditions to evaluate the constant of integration produces
\begin{equation}
\dot{x}=-2\omega_0y+\dot{x}_0+2\omega_0y_0.
\label{eq:tintegral}
\end{equation}
This is used to replace $\dot{x}$ in Eq.~(\ref{eq:CWy}).  The result is easily solved to yield, after the application of boundary conditions,
\begin{equation}
y(t)=-\left(3y_0+\dfrac{2\dot{x}_0}{\omega_0}\right)\cos(\omega_0t)+\dfrac{\dot{y}_0}{\omega_0}\sin(\omega_0t)+4y_0+\dfrac{2\dot{x}_0}{\omega_0}
\label{eq:yt}
\end{equation}
and
\begin{equation}
\dot{y}(t)=\left(3\omega_0y_0+2\dot{x}_0\right)\sin(\omega_0t)+\dot{y}_0\cos(\omega_0t).
\label{eq:ydott}
\end{equation} 
Substituting Eq.~(\ref{eq:ydott}) for $\dot{y}$ into Eq.~(\ref{eq:CWx}) gives
\begin{equation}
\ddot{x}=-\left(6\omega_0^2y_0+4\omega_0\dot{x}_0\right)\sin(\omega_0t)-2\omega_0\dot{y}_0\cos(\omega_0t).
\label{eq:xddot}
\end{equation}
Solving both the particular and homogeneous versions of Eq.~(\ref{eq:xddot}) and applying the initial conditions yields
\begin{equation}
x(t)= \left(6y_0+4\dfrac{\dot{x}_0}{\omega_0}\right)\sin(\omega_0t)+2\dfrac{\dot{y}_0}{\omega_0}\cos(\omega_0t)+x_0-2\dfrac{\dot{y}_0}{\omega_0}-\left(3\dot{x}_0+6\omega_0y_0\right)t
\label{eq:xt}
\end{equation}
and
\begin{equation}
\dot{x}(t)=\left(6\omega_0y_0+4\dot{x}_0\right)\cos(\omega_0t)-2\dot{y}_0\sin(\omega_0t)-3\dot{x}_0-6\omega_0y_0.
\label{eq:xdott}
\end{equation}
Equations~(\ref{eq:zt}), (\ref{eq:zdott}), (\ref{eq:yt}), (\ref{eq:ydott}), (\ref{eq:xt}), and (\ref{eq:xdott}) describe the motion of the interceptor seen from the target's reference frame $S$.  They are called the Clohessy-Wiltshire equations.\cite{CW}

\begin{figure}[h] 
  \centering
 \includegraphics[width=5.000in,height=2.6431in,keepaspectratio]{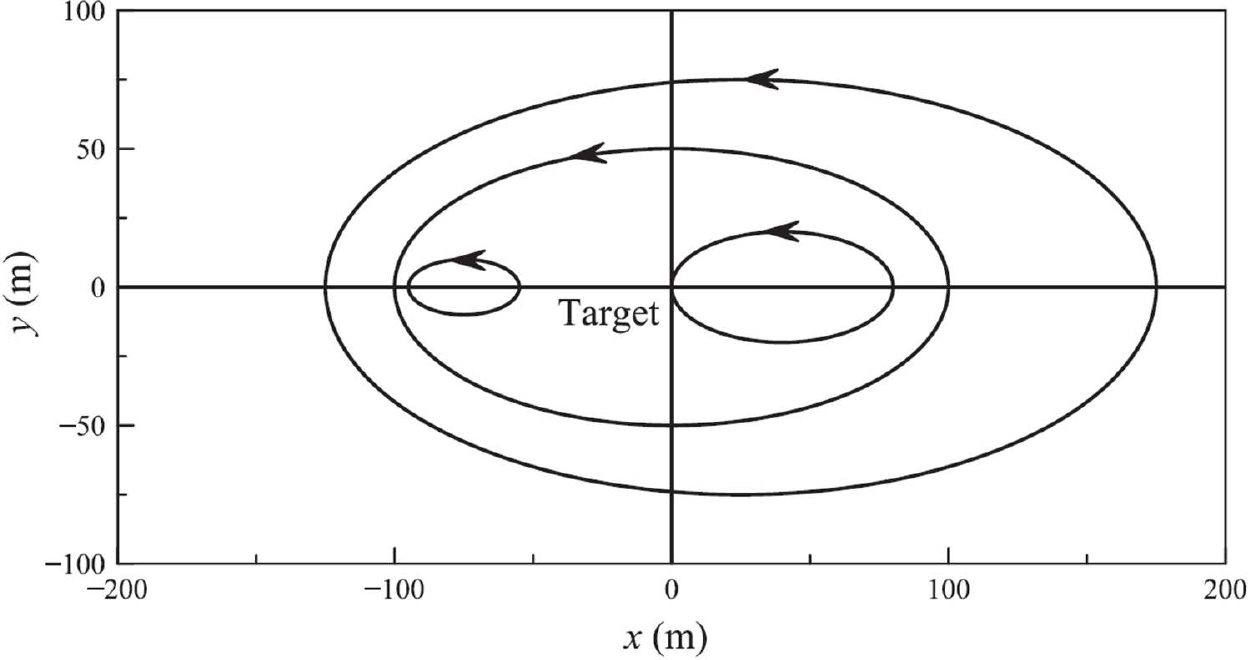}
  \caption{Several examples of stationary ellipses ($v_{\rm drift}=0$).  Note the ``boomerang'' ellipse; an astronaut aboard the target spacecraft could propel an object in the negative $y$-direction with such a speed that it would return to her after one orbital period.  The ``boomerang'' trajectory was also noted in Ref.~\onlinecite{Butikov}.}
\label{fig:ellipse}
\end{figure}

At this point, the usual procedure is to express the solutions for the positions and velocities as a single matrix equation, and then crank out the solutions as functions of time.\cite{Wiesel}  I want to follow and extend the analysis performed by Mueller in 1962 to uncover the implications of the Clohessy-Wiltshire equations.\cite{Mueller}  Let us continue to specify that $z(t)=\dot{z}(t)=0$, and just look at the motion in the $x$-$y$ plane.  For convenience, define 
\begin{eqnarray}
C&=&3y_0+2\dfrac{\dot{x}_0}{\omega_0}\label{eq:C}\\
D&=&\dfrac{\dot{y}_0}{\omega_0}\label{eq:D}\\
x_c&=&x_0-2\dfrac{\dot{y}_0}{\omega_0}-(3\dot{x}_0+6y_0\omega_0)t\label{eq:xc}\\
y_c&=&4y_0+2\dfrac{\dot{x}_0}{\omega_0}\label{eq:yc}
\end{eqnarray}
Equations~(\ref{eq:xt}) and (\ref{eq:yt}) are then
\begin{equation}
x-x_c=2C\sin(\omega_0t)+2D\cos(\omega_0t)
\label{eq:xmxc}
\end{equation}
and
\begin{equation}
y-y_c=-C\cos(\omega_0t)+D\sin(\omega_0t).
\label{eq:ymyc}
\end{equation}
Squaring Eqs.~(\ref{eq:xmxc}) and (\ref{eq:ymyc}) and then adding them leads to
\begin{equation}
\dfrac{(x-x_c)^2}{4(C^2+D^2)}+\dfrac{(y-y_c)^2}{C^2+D^2}=1.
\label{eq:ellipse}
\end{equation}
Formally, this is the equation of an ellipse with the center of the ellipse at the point $(x_c,y_c)$, $(x-x_c)^2/a^2+(y-y_c)^2/b^2=1$.  The semimajor axis $a$ of the ellipse is 
\begin{equation}
a=2\sqrt{C^2+D^2}=2\sqrt{\left(3y_0-2\dfrac{\dot{x}_0}{\omega_0}\right)^2+\left(\dfrac{\dot{y}_0}{\omega_0}\right)^2}
\label{eq:semimajora}
\end{equation} 
and the semiminor axis $b$ is half that,
\begin{equation}
b=\dfrac{a}{2},
\label{eq:semiminorb}
\end{equation}
which agrees with the specific example of Eqs.~(\ref{eq:examplex}) and (\ref{eq:exampley}). The eccentricity of the ellipse is $\sqrt{1-(b/a)^2}=\sqrt{3}/2$, independent of the initial conditions.  Of course, in general $x_c$ is a function of time, so the ellipse drifts parallel to the $x$-axis along the line $y=y_c$ with a drift velocity of
\begin{equation}
v_{\rm drift}=-(3\dot{x}_0+6\omega_0y_0)=-\dfrac{3}{2}\omega_0y_c.
\label{eq:vdrift}
\end{equation}
For a stationary ellipse, $v_{\rm drift}=0$ so $y_c=0$.  Let's find the initial conditions that will produce a stationary ellipse with specified values of the center $x_c$ and semimajor axis $a$.  Start by choosing (with no loss of generality) $y_0=\dot{x}_0=0$, so the initial conditions are for a point at the end of the major axis of the ellipse.  Equations~(\ref{eq:xc}) and (\ref{eq:semimajora}) then show that $\dot{y}_0=a\omega_0/2$ and $x_0=x_c+a$.  This stationary ellipse may be centered on the target ($x_c=0$), or may be arbitrarily offset along the $x$-axis ($x_c=x_0-2\dot{y}_0/\omega_0$).  Several examples are shown in Fig.~\ref{fig:ellipse}.

In light of Eq.~(\ref{eq:ellipse}) and Eqs.~(\ref{eq:C})--(\ref{eq:yc}), the specific example of Eqs.~(\ref{eq:examplex}) and (\ref{eq:exampley}) may be generalized to
\begin{equation}
x - x_c = a\cos(\omega_0t)
\label{eq:examplegenx}
\end{equation}
and
\begin{equation}
y - y_c = \frac{a}{2}\sin(\omega_ot).
\label{eq:examplegeny}
\end{equation}
As was done previously, these expressions can be used to calculate the Coriolis force, the ``tidal'' force, and the total force.  The result for the total force per unit mass on the interceptor is
\begin{equation}
\vec{F}_{\rm total}=\vec{F}_{\rm Cor}+\vec{F}_{\rm tidal}=-\omega_0^2\left[ (x - x_c)\,\hat{i}+(y - y_c)\,\hat{j}\right]=-\omega_0^2\left( \vec{r} - \vec{r}_c\right),
\label{eq:Ftotal}
\end{equation}      
where $\vec{r}_c$ is the position vector of the center of the elliptical trajectory, either drifting or stationary.  Thus $\vec{r} - \vec{r}_c$ is the vector from the center of the ellipse at time $t$ to the interceptor, and as before the total force on the interceptor obeys Hooke's law.  The interceptor acts as though it were attached to a spring, with the other end of the spring attached to the instantaneous center of the drifting or stationary ellipse. 

\begin{figure}[h] 
  \centering
  \includegraphics[width=7.000in,height=3.428in,keepaspectratio]{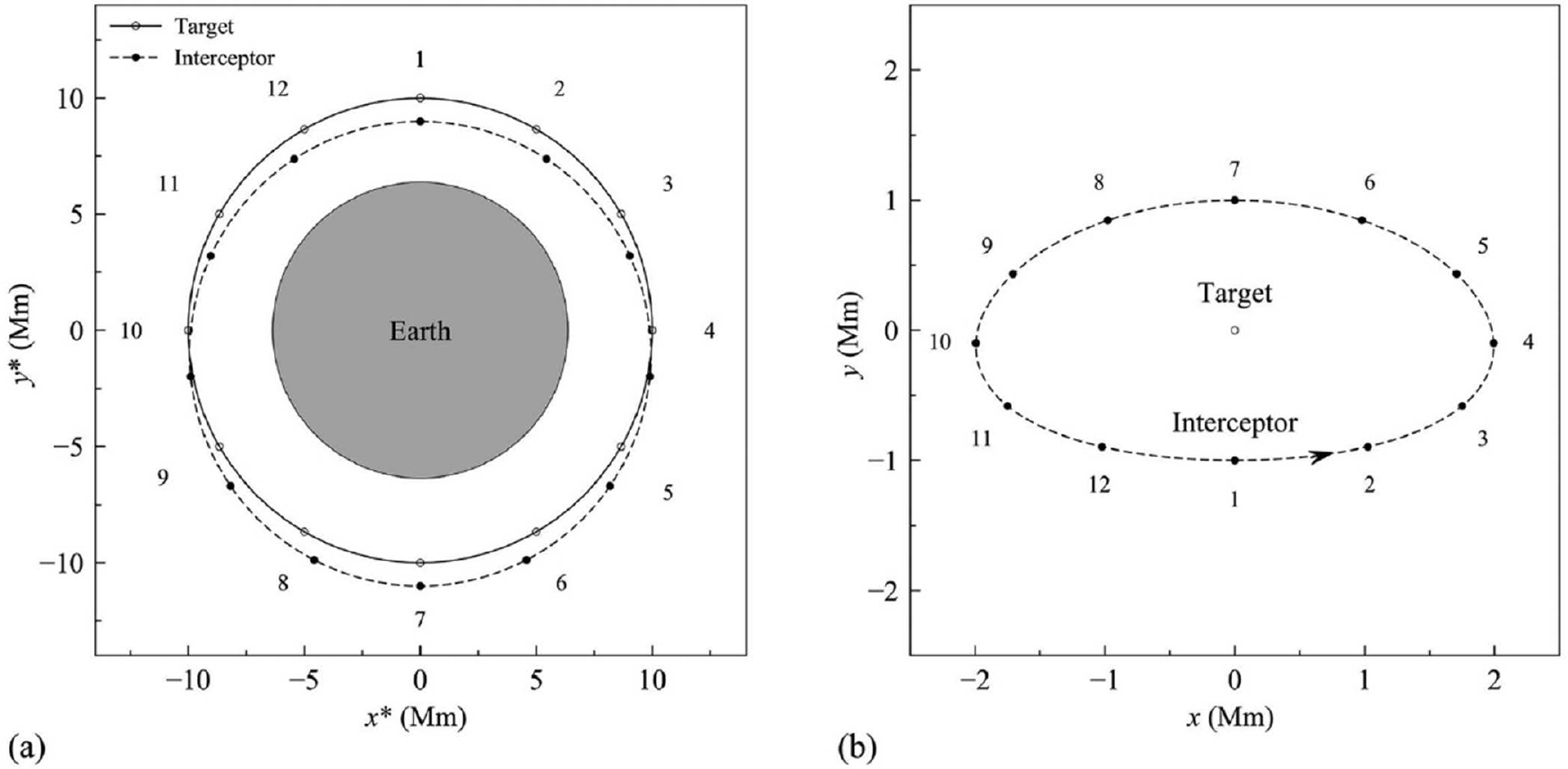}
  \caption{(a) As viewed in inertial frame $S^*$, the target occupies a circular orbit of radius $R_0=10^7$~m and the interceptor follows an elliptical orbit with semimajor axis $a^*=R_0$ and eccentricity 0.1.  The positions of the spacecraft were plotted at twelve equally spaced times using Kepler's equation.\cite{Prussing,Meeus}  (b) The motion of the interceptor as viewed in the non-inertial frame $S$ of the target with the $y$-axis directed radially outward.  The figure deviates from a perfect ellipse because the separation between the spacecraft is sufficiently large to violate the linearization requirement that $x$ and $y\ll R_0$.}
\label{fig:Kepler}
\end{figure}

How is the interceptor's motion about the target understood from the inertial frame $S^*$ of Earth?  As shown in Fig.~\ref{fig:Kepler}, the interceptor's stationary elliptical motion is caused by it being in a slightly non-circular (elliptical) orbit whose semimajor axis $a^*=R_0$.  At apogee and and perigee the interceptor will be carried outside and inside the target's orbit, respectively.  The alignment of the interceptor with the target and Earth's center at perigee (point 1) and apogee (point 7) was chosen to produce a qualitative agreement with Eqs.~(\ref{eq:examplex}) and (\ref{eq:exampley}).  In Fig.~\ref{fig:Kepler}, point 4 corresponds to $t = 0$ in these equations.  Other choices of alignment and orbital eccentricity (but with the same semimajor axis $a^*$) for the interceptor's orbit around Earth can reproduce the other stationary ellipses seen in Fig.~\ref{fig:ellipse}.

Returning to Eq.~(\ref{eq:vdrift}), if $y_c>0$ then the ellipse drifts in the negative $x$-direction, and if $y_c<0$, then the ellipse drifts in the positive $x$-direction.  In any case, in the rotating frame $S$ the interceptor travels in a counter-clockwise direction (seen from a point on the positive $z$-axis) around the ellipse.  In one orbital period, the center of the ellipse will be displaced by
\begin{equation}
\Delta x_c=v_{\rm drift}\left(\dfrac{2\pi}{\omega_0}\right)=-3\pi y_c.
\label{eq:Deltay}
\end{equation}

\begin{figure}[h] 
  \centering
  \includegraphics[width=5.67in,height=3.013in,keepaspectratio]{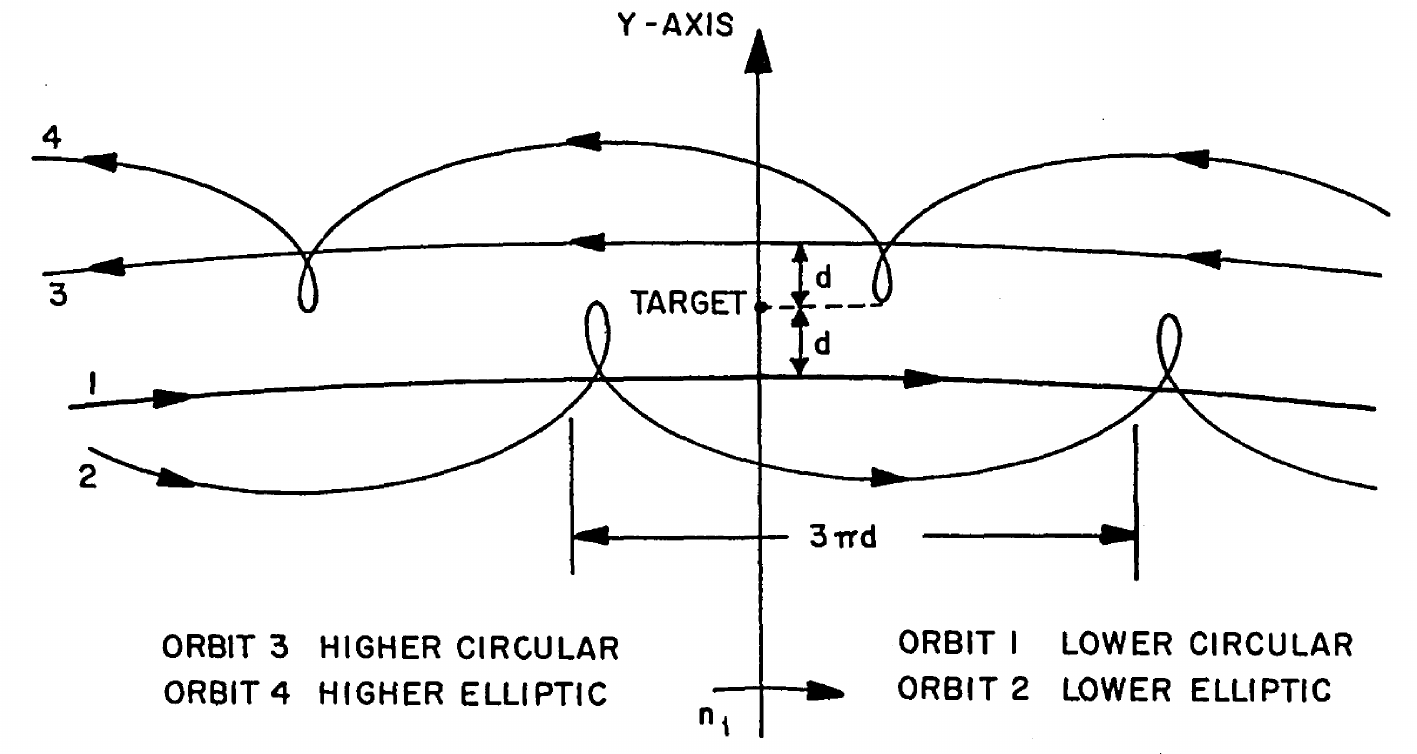}
  \caption{This figure is adapted from Buzz Aldrin's Ph.D.\ thesis and shows the interceptor's motions as seen from the target's reference frame $S$, with the target at $y=0$.  It has been flipped horizontally so the orientations of Aldrin's reference frames are the same as those in Fig~\ref{fig:axes}.  The arrow at the bottom shows that the positive $x$-axis is directed to the right, in the direction of the target's motion.  The distance $d$ in the vertical direction is the magnitude of $y_c$ in this paper.  The center of each ellipse drifts a distance of $3\pi d$ in one orbital period.}
\label{fig:Aldrin}
\end{figure}

A drifting ellipse occurs when the interceptor's elliptical orbit around Earth has $a^*>R_0$ ($y_c>0$) or $a^*<R_0$ ($y_c<0$).  By Kepler's third law (Eq.~\ref{eq:Kepler3rdgen}), when the semimajor axis of the interceptor's orbit has $a^*>R_0$ (or $a^*<R_0$), the orbital period is then greater than (or less than) the target's and the interceptor's apogee moves increasingly behind (or ahead of) the target.  Only if the semimajor axis $a^*=R_0$ will its orbital period is the same as the target's, so there is no drifting ($y_c=0$ so $v_{\rm drift}=0$).  

Figure~\ref{fig:Aldrin} from astronaut Buzz Aldrin's Ph.D.\ thesis shows the looping motions of two drifting ellipses.\cite{Aldrin}  Astronaut Michael Collins, exasperated by these loops as Gemini 10 attempted to dock with its target, called these loops ``wifferdills.''\cite{Collins}  Orbit~1 is the interceptor's circular orbit of radius $r^*<R_0$.  According to Kepler's 3rd law, if the interceptor is in a circular orbit with $r^* < R_0$, then its velocity is greater that the target's and so the interceptor appears to move in the  positive $x$-direction.  Orbit 2 is the interceptor's elliptical orbit of semimajor axis $a^*<R_0$.  According to Kepler's 3rd law, if the interceptor is in an elliptical orbit with $a^* < R_0$, then its orbital period is less than the target's.  For most of its orbit, the interceptor will therefore move faster than the target so the interceptor appears to move in the positive $x$-direction most of the time.  However, near apogee the interceptor can momentarily move more slowly than the target and so appear to move in the negative-$x$ direction, thereby producing Collins' ``wifferdill.''  The net result for orbit 2 is that the interceptor's trajectory, seen from a point on the the positive $z$-axis of the rotating target frame $S$, shows the interceptor moving with $y_c<0$ around a drifting ellipse in the counter-clockwise direction as the center of the ellipse travels in the positive $x$-direction.  Orbit~3 is the interceptor's circular orbit of radius $r^*>R_0$.  According to Kepler's 3rd law, if the interceptor is in a circular orbit with $r^* > R_0$, its velocity is less that the target's and so the interceptor appears to move in the  negative $x$-direction.  Orbit 4 is the interceptor's elliptical orbit of semimajor axis $a^*>R_0$.  According to Kepler's 3rd law, if the interceptor is in an elliptical orbit with $a^* > R_0$, then its orbital period is greater than the target's.  For most of its orbit, the interceptor will therefore move more slowly than the target so the interceptor appears to move in the negative $x$-direction most of the time.  However, near perigee the interceptor can momentarily move faster than the target and so appear to move in the positive-$x$ direction, again producing Collins' ``wifferdill.''  The net result for orbit 4 is that the interceptor's trajectory, seen from a point on the the positive $z$-axis of the rotating target frame $S$, shows the interceptor moving with $y_c>0$ around a drifting ellipse in the counter-clockwise direction as the center of the ellipse travels in the negative $x$-direction.  Aldrin's artfully worded dedication of his thesis is worth noting:  ``In the hopes that this work may in some way contribute to their exploration of space, this is dedicated to the crew members of this country's present and future manned space programs. If only I could join them in their exciting endeavors.''


\section{The Stranded Astronaut}\label{sec:astronaut}

\begin{figure}[h] 
  \centering
  \includegraphics[width=6.5in,height=1.39in,keepaspectratio]{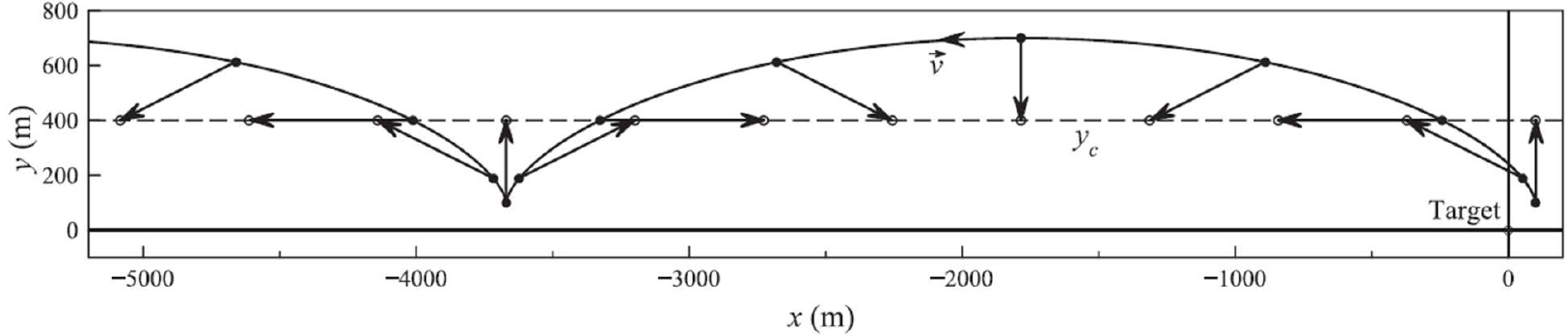}
  \caption{The trajectory of an astronaut who starts at rest at $x_0=y_0=100$~m.  The solid dots show the position of the astronaut after 12 equal time intervals, and the 12 open circles along the horizontal line $y = y_c$ show the center of the drifting ellipse at those same times.  The vectors from the astronaut to the ellipse centers are the total force acting on the astronaut, which obeys Hooke's law, Eq.~(\ref{eq:Ftotal}).  Imagining these vectors as an extended spring provides a visceral feel for the total force acting on the astronaut.}
\label{fig:atRest}
\end{figure}

Suppose an astronaut (the interceptor) is stranded at some distance from her spacecraft (the target).  Specifically, let us suppose that her spacecraft is in orbit at the same altitude above Earth as the International Space Station (400~km), where $\omega_0=1.13\times 10^{-3}$~rad/s, and the orbital period is 92.4~min.  Let her be initially at rest in frame $S$ at $x_0=y_0=100$~m.  If she does nothing, the ``tidal'' force will exert an upward force on her.  The Coriolis force will then bend her trajectory as shown in Fig.~\ref{fig:atRest}, carrying her increasingly far from the target.  Unless $y_0=0$, this ``hopping'' path will occur regardless of her starting point because if $\dot{x}_0$ and $\dot{y}_0$ are zero, then Eqs.~(\ref{eq:xt}) and (\ref{eq:yt}) become
\begin{eqnarray}
x(t)-x_0&=&-6y_0[\omega_0t-\sin(\omega_0t)] \label{eq:xthop}\\
y(t)-y_0&=&3y_0[1-\cos(\omega_0t)] \label{eq:sthop}
\end{eqnarray}
These are similar to the parametric equations for a cycloid, except here the magnitudes of the leading coefficients on the right-hand side are not equal.  However, this is indeed a drifting ellipse with drift speed $v_{\rm drift}=-a\omega_0$, with semimajor axis $a=6y_0$.  The ``tidal'' and Coriolis forces combine to produce a Hooke's law force in the direction of the center of the drifting ellipse, as shown in in the figure.

\begin{figure}[h] 
  \centering
  \includegraphics[width=4.0in,height=3.845in,keepaspectratio]{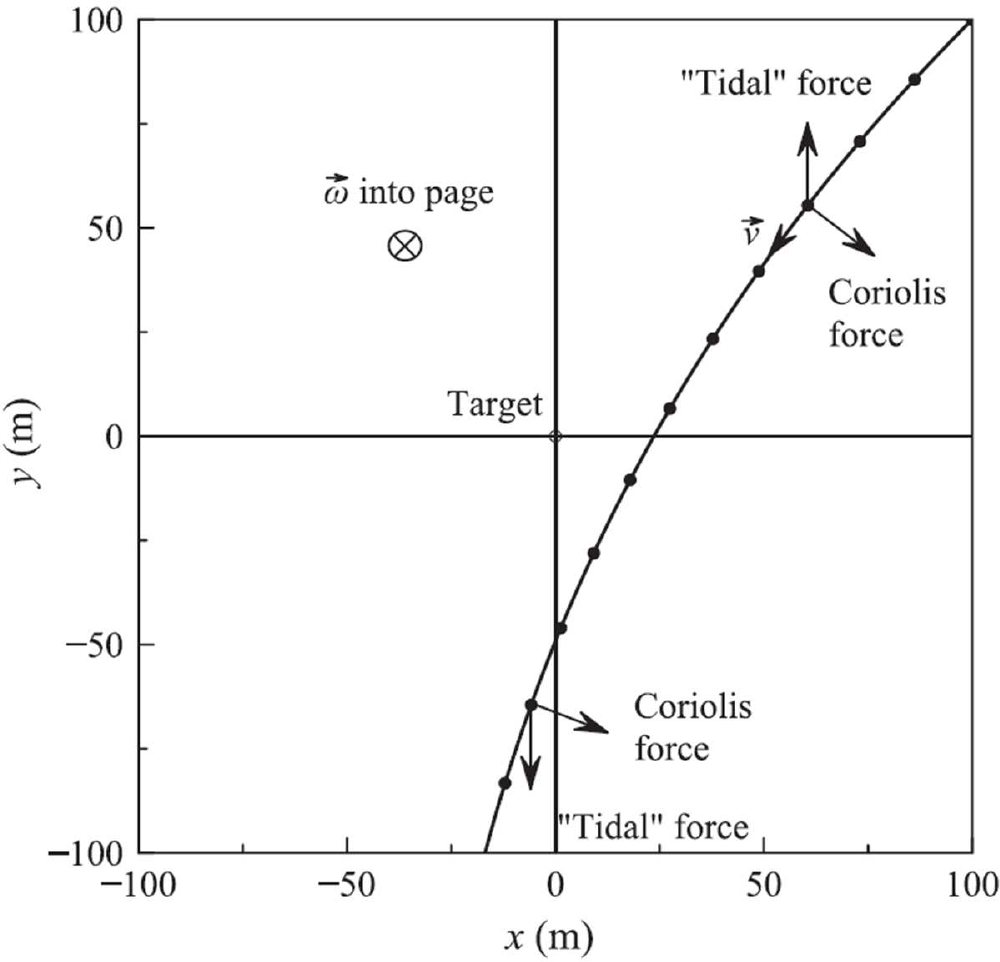}
  \caption{The trajectory of an astronaut attempting to return to her target spacecraft from an initial position of $x_0=100$~m and $y_0=100$~m.  The dots mark her position at 20-s intervals.  The astronaut's velocity $\vec{v}=d\vec{r}/dt$, is shown, along with the directions of the ``tidal'' force and the Coriolis force which determine her trajectory. At these two points near the $x$-axis, the magnitude of the ``tidal force'' is about 10 percent that of the Coriolis force.  The direction of the angular velocity vector $\vec{\omega}$ is also indicated.  The dominant role of the Coriolis force in this situation was also described qualitatively in Ref.~\onlinecite{Michalik}.}  
\label{fig:astronaut01}
\end{figure}

\begin{figure}[h] 
  \centering
  \includegraphics[width=6.5in,height=1.46in,keepaspectratio]{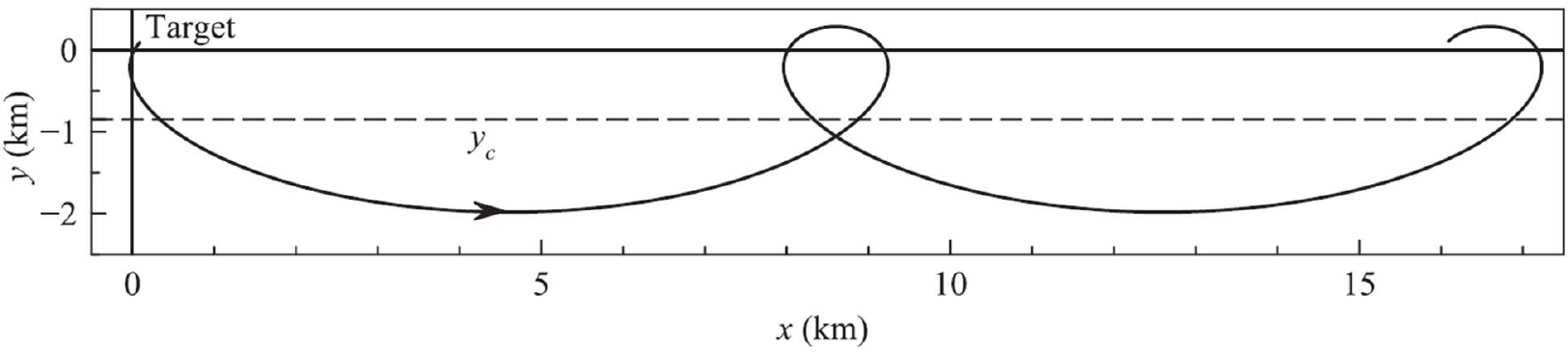}
  \caption{The trajectory of an astronaut attempting to return to her target spacecraft from an initial position of $x_0=100$~m and $y_0=100$~m during two orbital periods.  Her trajectory carries her away from the target as the center of her drifting elliptical path travels along the line $y=y_c$.}
\label{fig:astronaut02}
\end{figure}

Fortunately the astronaut has a small thruster unit that can produce a change in speed of $\Delta v=1$~m/s.  If she has enough fuel for a single thrust, what should she do to return to the safety of her spacecraft?  The obvious solution requires that she produce a thrust directly toward the spacecraft, so the components of her initial velocity are $\dot{x}_0=\dot{y}_0=-1/\sqrt{2}$~m/s.  Alas, as Fig.~\ref{fig:astronaut01} shows, as soon she begins moving toward the spacecraft, the Coriolis and ``tidal'' forces deflect her trajectory away from the straight line she intends, bending it downward below the spacecraft into a counter-clockwise arc.  Her distance of closest approach is 20.8~m as she passes below her target.  Her motion continues to slowly loop in a drifting ellipse whose center moves along the line $y=y_c=-0.848$~km at a velocity of $v_{\rm drift}=1.44$~m/s.  In one orbital period the center of her ellipse is displaced by $\Delta x_c=-3\pi y_c=7.99$~km.  Figure~\ref{fig:astronaut02} shows the astronaut's trajectory over two orbital periods.

\begin{figure}[h] 
  \centering
  \includegraphics[width=4.0in,height=3.089in,keepaspectratio]{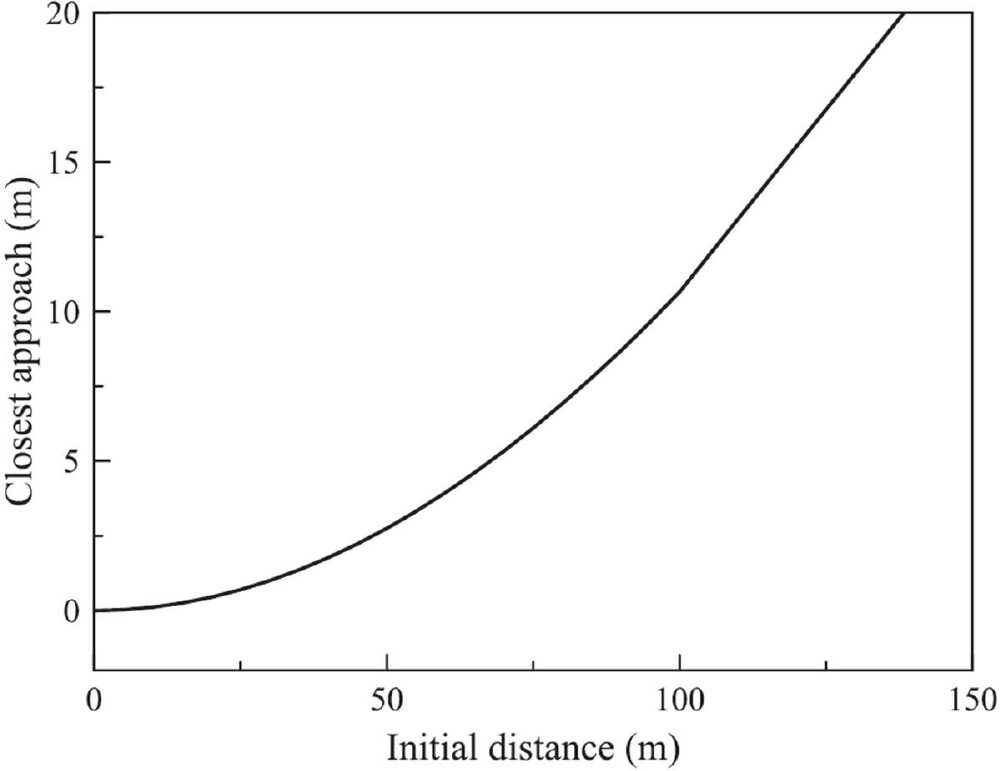}
  \caption{The distance of closest approach of an astronaut to the target, starting at $x_0=y_0$.}
\label{fig:closest}
\end{figure}
 
Figure~\ref{fig:closest} shows the results of moving the astronaut's initial position toward the target along the line $x_0=y_0$.  She is initially at rest before the thruster fires.  Numerically, it is found that her distance of closest approach is proportional to the square of the her initial distance.  Assume the stranded astronaut will successfully return to her spacecraft if it is within her reach.  For an initial distance of 30~m, her distance of closest approach is 1.00 m, and for an initial distance of 40~m (Shirra's quoted distance) it is 1.77~m, the height of a 5~ft 10~in astronaut, so this is the critical range that separates a successful from an unsuccessful rendezvous attempt for our astronaut.  [Equation~(4a) of Ref.~\onlinecite{Freedman} differs from this paper's Eq.~(\ref{eq:CWy}) (our $x$'s and $y$'s are reversed), and so our results for a stranded astronaut are difficult to compare.]

All of the guesswork can be removed by setting the left-hand sides of Eqs.~(\ref{eq:yt}) and (\ref{eq:xt}) equal to zero and then using specified values of $x_0$, $y_0$, and an assumed time-of-flight $t_f$ to solve them for the initial values of $\dot{x}_0$ and $\dot{y}_0$.  The results are
\begin{equation}
\dot{x}_0=\dfrac{\omega_0x_0\sin(\omega_0t_f)-\omega_0y_0\{6\omega_0t_f\sin(\omega_0t_f)-14[1-\cos(\omega_0t_f)]\}}{3\omega_0t_f\sin(\omega_0t_f)-8[1-\cos(\omega_0t_f)]}
\label{eq:xdot0}
\end{equation}
and
\begin{equation}
\dot{y}_0=\dfrac{-2\omega_0x_0[1-\cos(\omega_0t_f)]+\omega_0y_0[4\sin(\omega_0t_f)-3\omega_0t_f\cos(\omega_0t_f)]}{3\omega_0t_f\sin(\omega_0t_f)-8[1-\cos(\omega_0t_f)]}.
\label{eq:ydot0}
\end{equation}  

The interceptor may already have a pre-thrust velocity, $\vec{v}_{\rm pre}$, so to reach the needed values of $\dot{x}_0$ and $\dot{y}_0$ only a {\em velocity increment} $\Delta\vec{v}$ needs to be added to the pre-thrust velocity.  Let us define the {\em aiming angle} $\theta_{\rm aim}$ for a successful rendezvous with the target as the direction of this velocity increment, 
\begin{equation}
\tan(\theta_{\rm aim})=\dfrac{\Delta v_y}{\Delta v_x}=\dfrac{\dot{y}_0-v_{{\rm pre},y}}{\dot{x}_0-v_{{\rm pre},x}}.
\label{eq:thetaaim}
\end{equation}
Subtracting the pre-thrust velocity is not necessarily a negligible correction, as will be seen in Sec.~\ref{sec:Apollo11} for the Apollo~11 rendezvous above the Moon.

\begin{figure}[h] 
  \centering
  \includegraphics[width=3.921in,height=3.0in,keepaspectratio]{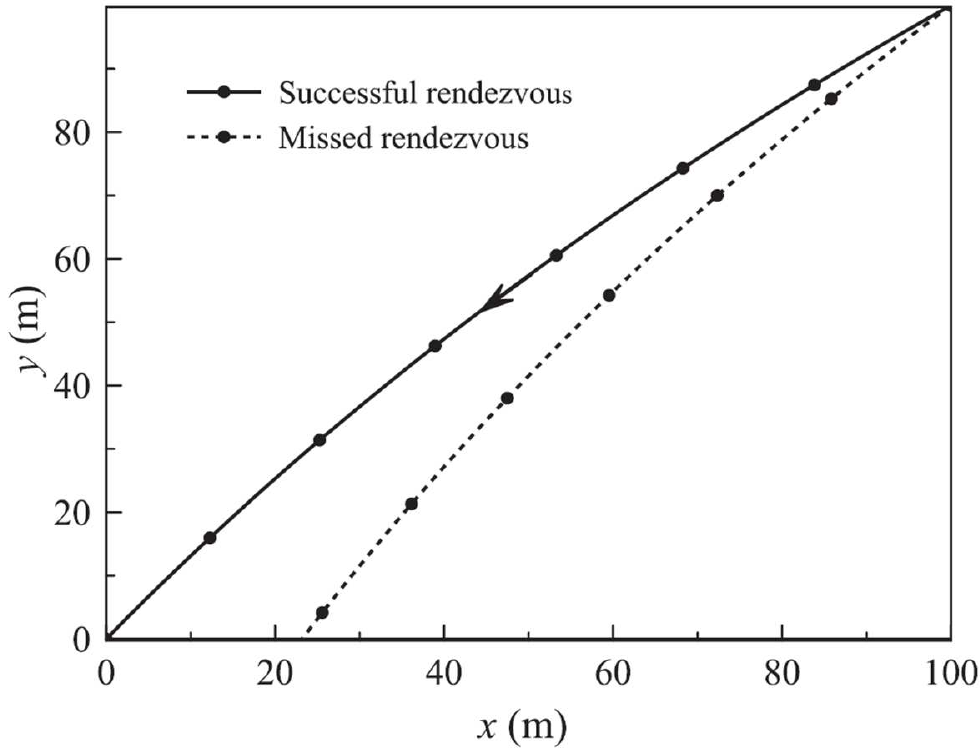}
  \caption{The astronaut's successful and missed rendezvous with her target spacecraft from an initial position of $x_0=100$~m and $y_0=100$~m.  The dots mark her position at 20-s intervals.  The initial speeds for the two rendezvous attempts are the same.}
\label{fig:success}
\end{figure}

Now return to the stranded astronaut, at rest at $x_0=y_0=100$~m before her thruster fires.  Her distance from the target is $100\sqrt{2}$~m and she wants to arrive with a velocity of about 1~m/s, so she chooses a time-of-flight of $t_f=140$~s.  Inserting these values into Eqs.~(\ref{eq:xdot0}) and (\ref{eq:ydot0}), the components of her initial velocity are $\dot{x}_0=-0.822$~m/s and $\dot{y}_0=-0.614$~m/s, so her initial speed should be 1.026~m/s.  Her aiming angle is, from Eq.~(\ref{eq:thetaaim}), $\theta_{\rm aim}=216.7^\circ$.  This is $36.7^\circ$ below the negative $x$-axis, and above the direction of her earlier attempt.  Equations~(\ref{eq:ydott}) and (\ref{eq:xdott}) reveal that she arrives at her target with a gentle speed of 1.01~m/s.  Figure~\ref{fig:success} shows the astronaut's successful trajectory back to the target spacecraft, along with her missed trajectory for comparison.  (The missed trajectory was recalculated so the initial speeds were the same.) 

\section{Line-of-Sight Targeting}\label{sec:los}

\begin{figure}[h] 
  \centering
  \includegraphics[width=6.5in,height=2.644in,keepaspectratio]{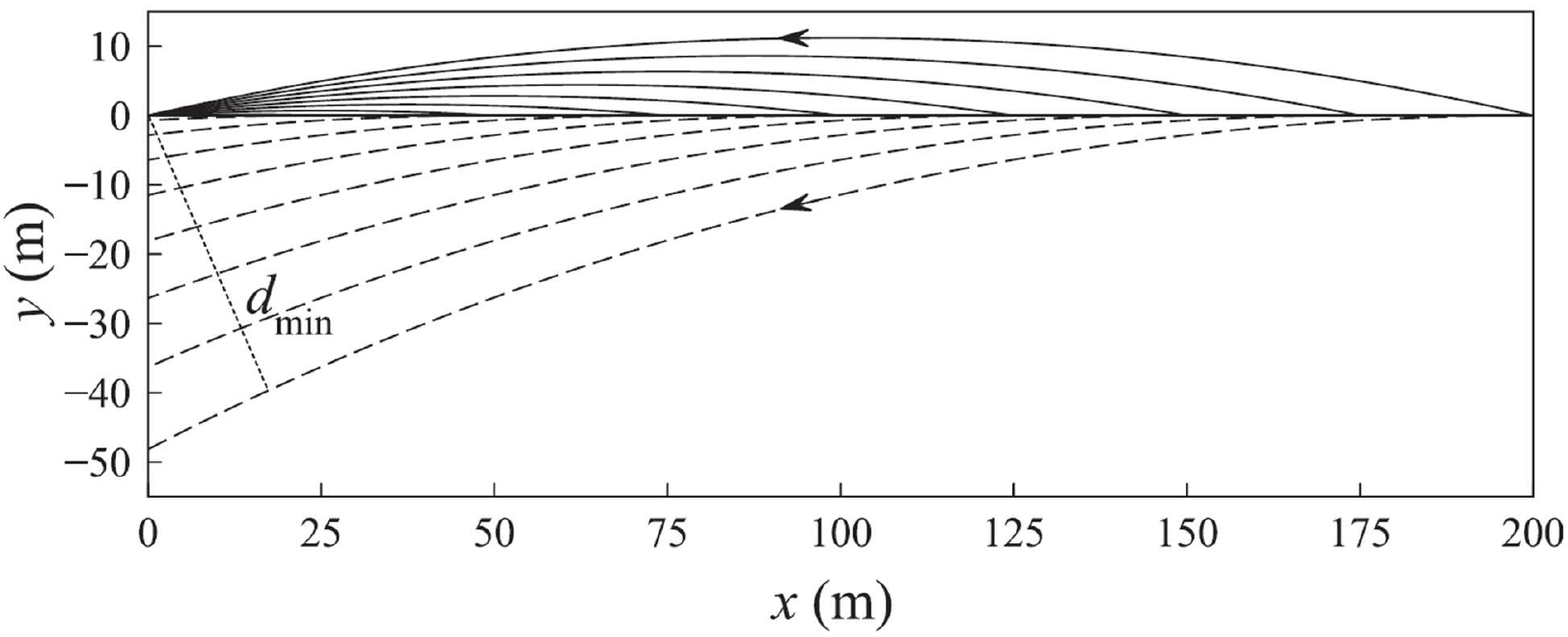}
  \caption{Trajectories for $x_0 = $ multiples of 25~m.  The upper trajectories use Eqs.~(\ref{eq:xdot0}) and (\ref{eq:ydot0}) and times-of-flight $t_f=x_0/(1$~m/s).  The lower trajectories use line-of-sight targeting.  The target is at the origin, and the distance of closest approach $d_{\rm min}$ is shown for the lowest trajectory.}
\label{fig:xaxis}
\end{figure}

Recall Wally Schirra's statement that ``I don't think rendezvous is over until you are stopped -- completely stopped -- with no relative motion between the two vehicles, at a range of approximately 120 feet [about 40 meters].  That's rendezvous! From there on, it's stationkeeping. That's when you can go back and play the game of driving a car or driving an airplane or pushing a skateboard -- it's about that simple.''  Stationkeeping relies on the use of thrusters to maintain the interceptor spacecraft at rest relative to the target.  The only way the two vehicles can remain at relative rest without the continual firing of thrusters is if the two spacecraft occupy the same orbit, so in this section I will assume the interceptor is initially at rest at $y_0=0$.  Equations~(\ref{eq:xdot0}) and (\ref{eq:ydot0}), along with an assumed time-of-flight $t_f$, provide values of the initial $x$- and $y$-components of the velocity that will lead to a successful rendezvous with the target.  The trajectories to rendezvous from several initial positions ahead of the target are shown in the upper part ($y>0$) of Fig.~\ref{fig:xaxis}.  For this figure, $t_f=x_0/(1~\mbox{m/s})$.  The conservation of energy [Eq.~(\ref{eq:Em})] assures us that the speed of the interceptor is nearly constant along these shallow trajectories.

In the lower part of Fig.~\ref{fig:xaxis} are trajectories from the same initial positions as used in the upper part, and these trajectories use the same initial velocities as their counterparts in the upper part.  However, line-of-sight targeting is employed, with the initial velocity aimed directly toward the target (so $|\dot{x}_0|=1$~m/s and $\dot{y}_0=0$).  The figure illustrates the minimum distance, $d_{\rm min}$, by which the interceptor misses the target.  Let us approximate $d_{\rm min}$ as the $y$-coordinate of the intersection of a dashed trajectory in Fig.~\ref{fig:xaxis} with the $y$-axis, which occurs at $t\simeq |x_0/\dot{x}_0|$. 

Substituting $y_0=0$ and $\dot{y}_0=0$ into Eq.~(\ref{eq:yt}) produces
\begin{equation}
y(t)=\frac{2\dot{x}_0}{\omega_0}[1-\cos(\omega_0t)].
\end{equation}
The time of flight is on the order of 100~s, so $\omega_0 t\ll 1$.  Using the first two terms in a Taylor series for $\cos(\omega_0t)$ gives
\begin{equation}
y(t)\simeq \frac{2\dot{x}_0}{\omega_0}\,\frac{(\omega_0t)^2}{2}.
\end{equation}  
Finally, set $y(t) \simeq d_{\rm min}$ when $t = |x_0/\dot{x}_0|$ to obtain
\begin{equation}
d_{\rm min}\simeq \dfrac{\omega_0x_0^2}{|\dot{x}_0|}.
\label{eq:lineofsight}
\end{equation}
Although not exact, this relation is accurate to within 5~percent of the actual distance of closest approach in Fig.~\ref{fig:xaxis} when $x_0\leq 235$~m.  Note that the distance of closest approach is proportional to the square of the initial distance, as was also found for the stranded astronaut in Sec.~\ref{sec:astronaut}.  However, Eq.~(\ref{eq:lineofsight}) is not accurate if $y_0\neq 0$.  

Shirra spoke of ``driving a car'' to rendezvous.  The U.S.\ Interstate Highway System uses a 12~ft (3.66~m) standard lane width.  Let us require for a successful rendezvous that the car's center remain in its lane; that is, the distance by which the interceptor misses the target must satisfy
\begin{equation}
d_{\rm min} = \frac{\omega_0x_0^2}{|\dot{x}_0|}\leq 1.83~\text{m}.
\end{equation}
As before, $\omega_0=1.13\times 10^{-3}$~rad/s and $|\dot{x}_0| = 1$~m/s, so solving for $x_0$ results in a line-of-sight rendezvous range of $x_0\leq 40.24$~m.  Just as the astronaut said, within about 40~m the interceptor can aim directly at the target and fire its thrusters to arrive within 2~m of it.   

\section{Apollo 11's Lunar Rendezvous}\label{sec:Apollo11}

\begin{figure}[h] 
  \centering
  \includegraphics[width=6.5in,height=3.999in,keepaspectratio]{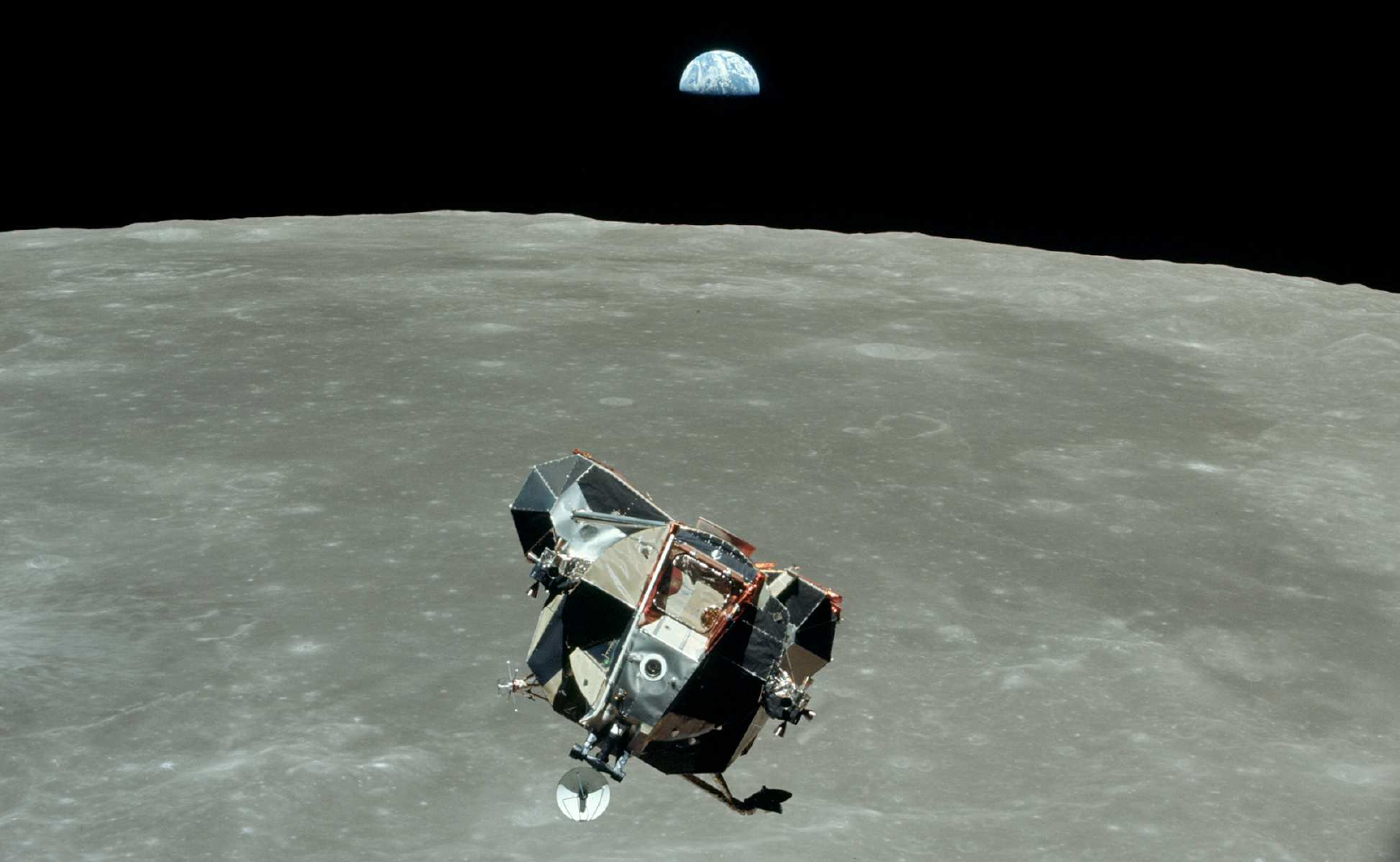}
  \caption{The rendezvous of the Apollo 11 lunar module (LM) with the command/service module (CSM).  This is known as ``The Loneliest Man'' because the photographer, Michael Collins in the CSM, was the only person in all of Earth's history not ``in'' the photo.  The spacecraft are separated by about 30~m, close enough for line-of-sight targeting. An RCS thruster unit, each with four nozzles, can be seen at the middle of the right and left sides of the LM.  Photo courtesy of NASA.}
\label{fig:rendezvous}
\end{figure}  

\begin{figure}[h] 
  \centering
  \includegraphics[width=5.265in,height=3.50in,keepaspectratio]{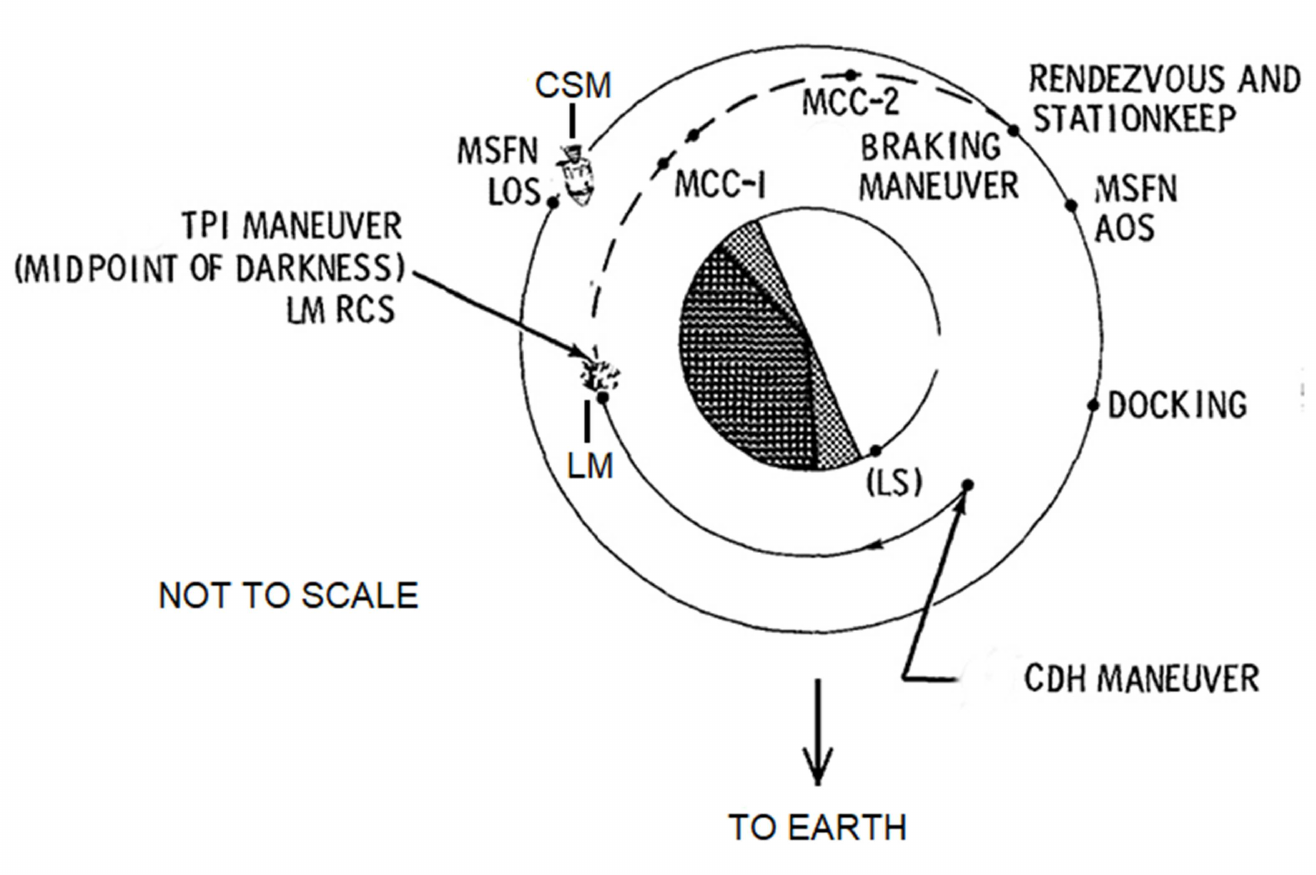}
  \caption{The final stages of Apollo 11's rendezvous of the lunar module (LM) with the command/service module (CSM); the earlier ascent from the LS (Landing Site) is not shown.  ``MSFN'' stands for ``Manned Space Flight Network,'' and ``LOS'' and ``AOS refer to the loss and later acquisition of the CSM's radio communications with Earth as the CSM passed behind the Moon.  The CDH maneuver is a firing of the RCS thrusters to achieve an orbit at a Constant Delta Height below the orbit of the CSM.  The later TPI (Terminal Phase Initiation) firing of the RCS thrusters is the subject of this section.  The two MCC's (Mid-Course Corrections) are not included in this analysis.  The figure is not to scale.  Actually, the height of the CSM above the surface was only 6.4 percent of the Moon's radius.  This figure was adapted from one in Ref.~\onlinecite{Journal}.}
\label{fig:TPI}
\end{figure}

\begin{figure}[h] 
  \centering
  \includegraphics[width=5.654in,height=3.0in,keepaspectratio]{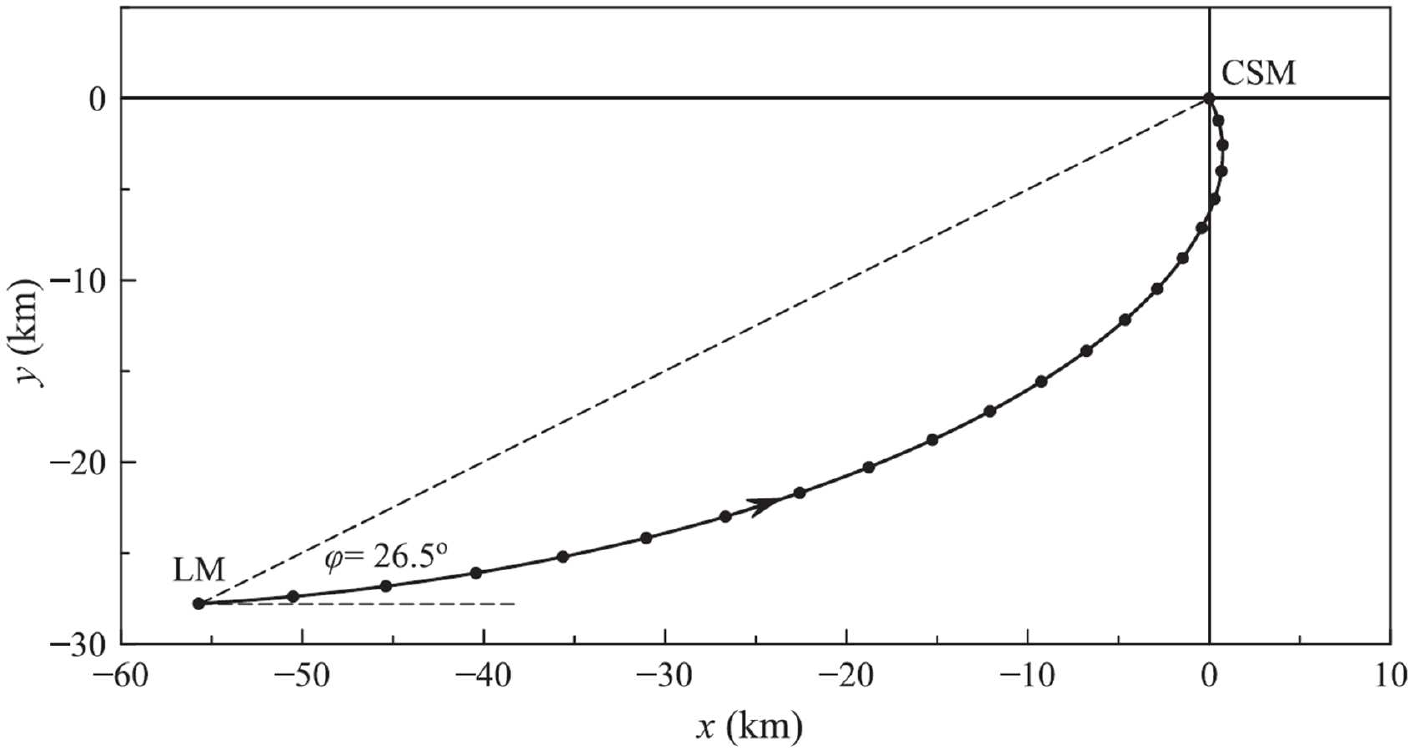}
  \caption{The rendezvous of the lunar module (LM) with the command/service module (CSM) of Apollo 11.  The dots mark the LM's position at 2-min intervals.}
\label{fig:Apollo}
\end{figure}

On July 21, 1969 (UT), while Neil Armstrong and Buzz Aldrin were planting the first footsteps on the Moon, Michael Collins piloted the command/service module (CSM) far above them.  Fifteen hours later, the ascent stage of the lunar module (LM; the interceptor) lifted off from the Sea of Tranquility to rendezvous with the CSM (the target).  The CSM was in a slightly elliptical orbit of 56.6 x 62.5 nautical miles,\cite{Journal2} but the following analysis will adopt a circular orbit at an altitude of 60.0~nautical miles (111.12~km) above the lunar surface.  At this altitude the CSM's angular velocity was 
\begin{equation}
\omega_0=\sqrt{\dfrac{GM}{R_0^3}}=8.81\times 10^{-4}\mbox{\ rad/s},
\label{omegaCMS}
\end{equation}
corresponding to an orbital period of 118.81~min.  Three burns of the LM's Reaction Control System (RCS) thrusters then placed the LM into a circular lunar orbit in the same plane as the CSM but $h = 15.0$~nautical miles (27.78~km) below it.  When the CSM appeared at an elevation angle of $\phi = 26.5^\circ$ above the local horizon of the LM (in the direction of motion; see Fig.~\ref{fig:Apollo}), the RCS thrusters fired for the Terminal Phase Initiation (TPI) maneuver, sending the LM coasting to rendezvous with the CSM 42~min later; see Fig.~\ref{fig:rendezvous}.\cite{NASA}  Although this description of the rendezvous process is simplified -- two midcourse corrections are ignored, as well as the later braking and stationkeeping maneuvers that finally led to a successful rendezvous and docking -- the calculated trajectory still shows the essential features of the TPI maneuver; see Fig.~\ref{fig:TPI}.\cite{Apollo11}

The initial position of the LM is $y_0=-h=-27.78$~km and $x_0=y_0\cot(26.5^\circ)=-55.72$~km.  These values, along with $t_f=42$~min, can be inserted into Eqs.~(\ref{eq:xdot0}) and (\ref{eq:ydot0}) to obtain the initial velocity components $\dot{x}_0=43.73$~m/s and $\dot{y}_0=2.53$~m/s.  Eqs.~(\ref{eq:yt}) and (\ref{eq:xt}) are then used to determine the lunar module's trajectory, as shown in Fig.~\ref{fig:Apollo}.  The dots in the figure are at 2-min intervals.  The LM's trajectory takes it slightly in front of the CSM, and during the final minutes the LM approaches the CSM along a gently curving track so its angular position, seen from the CSM, varies slowly.  The LM arrives with a speed of 10.9~m/s.  Several braking maneuvers then bring the LM to a stationkeeping position as discussed in Sec.~\ref{sec:los}.  The final moments of the TPI maneuver, followed by the beginning of the stationkeeping that led to the final docking of the spacecraft, can be seen in the video ``Apollo 11 - 16-mm magazine 1122-D.''\cite{Apollovideo}  The CSM's camera is looking down at the Moon in roughly the negative $y$-direction as the CSM travels in the positive $x$-direction toward the limb of the Moon in the bottom-right corner of the video.  

\begin{figure}[h] 
  \centering
  \includegraphics[width=2.49in,height=3.0in,keepaspectratio]{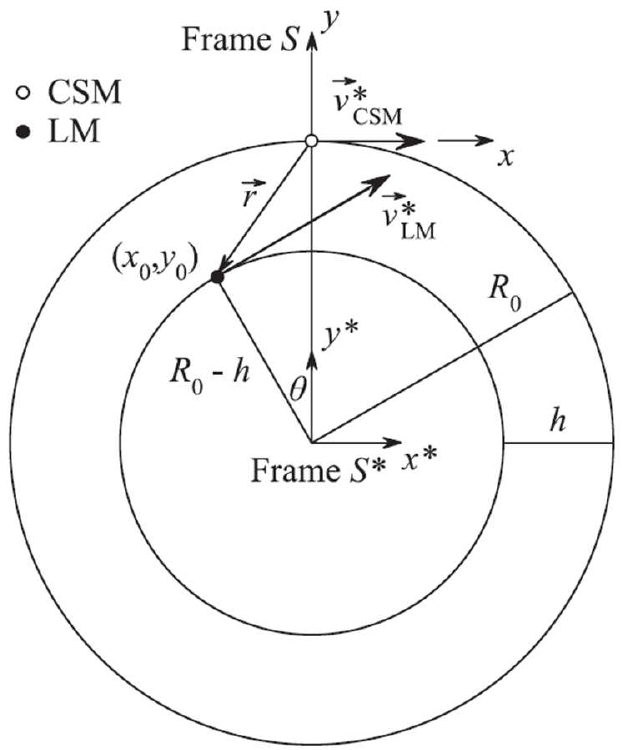}
  \caption{The coordinate systems used to calculate the pre-thrust velocity of the LM as seen from the non-inertial frame $S$.  The $y$ and $y^*$ axes are chosen to point in the same direction without loss of generality.  For clarity the Moon is not shown.  The orbits are not to scale.}
  \label{fig:vi}
\end{figure}

Before its thrusters were fired for the TPI maneuver, the LM was in a circular orbit at a distance $h$ below the CSM.  Because it was already moving relative to the CSM, the $x$ and $y$-components of its pre-thrust velocity must be subtracted from $\dot{x}_0$ and $\dot{y}_0$ to calculate the change in the velocity, $\Delta \vec{v}$, of the LM required for rendezvous.  Equation~(\ref{eq:v}) can be used to calculate the pre-thrust velocity of the LM in the rotating frame $S$ of the CSM.  
\begin{equation}
\vec{v}_{\rm pre}=\dfrac{d\vec{r}}{dt}=\dfrac{d^*\vec{r^*}}{dt}-\dfrac{d^*\vec{R^*}}{dt}-\vec{\omega}\times \vec{r}.
\label{eq:vi1}
\end{equation}
In Fig.~\ref{fig:vi}, the origin of the non-inertial frame $S$ (the CSM) is located on the $y^*$-axis a distance $R_0$ above the origin of the inertial frame $S^*$ (the center of the Moon).  The velocity of the CSM measured in the inertial frame $S^*$ is
\begin{equation}
\vec{v}^*_{\rm CSM}=\dfrac{d^*\vec{R}^*}{dt}=\omega_0R_0\hat{i},
\label{eq:vCSM}
\end{equation}
and the velocity of the LM measured in the inertial frame $S^*$ is
\begin{equation}
\vec{v}^*_{\rm LM}=\dfrac{d^*\vec{r}^*}{dt}=\omega_{\rm LM}(R_0-h)[\cos(\theta)\hat{i}+\sin(\theta)\hat{j}],
\label{eq:vLM}
\end{equation}
where $\omega_{\rm LM}$ is the angular velocity of the LM,
\begin{equation}
\omega_{\rm LM}=\sqrt{\dfrac{GM}{(R_0-h)^3}}=9.02\times 10^{-4}\mbox{\ rad/s}.
\label{eq:omegaLM}
\end{equation}
Keeping in mind that, in Fig.~\ref{fig:vi}, $x_0=x^*_0$ and $y_0=y^*_0-R_0$, some simple geometry leads to
\begin{equation}
\cos(\theta)=\dfrac{R_0+y_0}{R_0-h}
\label{eq:cos}
\end{equation}
and
\begin{equation}
\sin(\theta)=\dfrac{-x_0}{R_0-h}
\label{eq:sin}
\end{equation} 
Also, the position of the LM as seen from frame $S$ is
\begin{equation}
\vec{r}=x_0\hat{i}+y_0\hat{j}.
\label{eq:rLM}
\end{equation}
and $\vec{\omega}=-\omega_0\hat{k}$, so
\begin{equation}
\vec{\omega}\times \vec{r}=\omega_0y_0\hat{i}-\omega_0x_0\hat{j}.
\label{eq:omegacrossr}
\end{equation}
Inserting these into Eq.~(\ref{eq:vi1}) shows that the pre-thrust velocity of the LM as measured from frame $S$ is 
\begin{equation}
\vec{v}_{\rm pre}=(\omega_{\rm LM}-\omega_0)[(R_0+y_0)\hat{i}-x_0\hat{j}],
\label{eq:vi2}
\end{equation}
This expression is exact.  Assuming that $x_0$ and $y_0\ll R_0$, $R_0-h$ in Eq.~(\ref{eq:omegaLM}) can be replaced by $R_0+y_0$ and the components of the pre-thrust velocity are, to first order in smallness,
\begin{equation}
v_{{\rm pre},x}\simeq -\dfrac{3}{2}\omega_0y_0=36.73\mbox{\ m/s}
\label{eq:vix}
\end{equation}
and 
\begin{equation}
v_{{\rm pre},y}\simeq 0.
\label{eq:viy}
\end{equation}
The components of the change in the LM's velocity required for rendezvous are therefore given by
\begin{equation}
\Delta v_x=\dot{x}_0-v_{{\rm pre},x}=7.00\mbox{\ m/s}
\label{eq:Deltavx}
\end{equation}
and 
\begin{equation}
\Delta v_y=\dot{y}_0-v_{{\rm pre},y}=2.53\mbox{\ m/s}.
\label{eq:Deltavx}
\end{equation}
The magnitude of $\Delta \vec{v}$ is 7.44~m/s.  This is in good agreement with the nominal value of $\Delta v=7.56$~m/s called for in the Apollo 11 flight plan.  The aiming angle is [Eq.~(\ref{eq:thetaaim})]
\begin{equation}
\tan{\theta_{\rm aim}}=\dfrac{\Delta v_y}{\Delta v_x},
\label{eq:thetaaim2}
\end{equation}
so $\theta_{\rm aim}=19.8^\circ$.

\begin{figure}[h] 
  \centering
  \includegraphics[width=5.518in,height=3.0in,keepaspectratio]{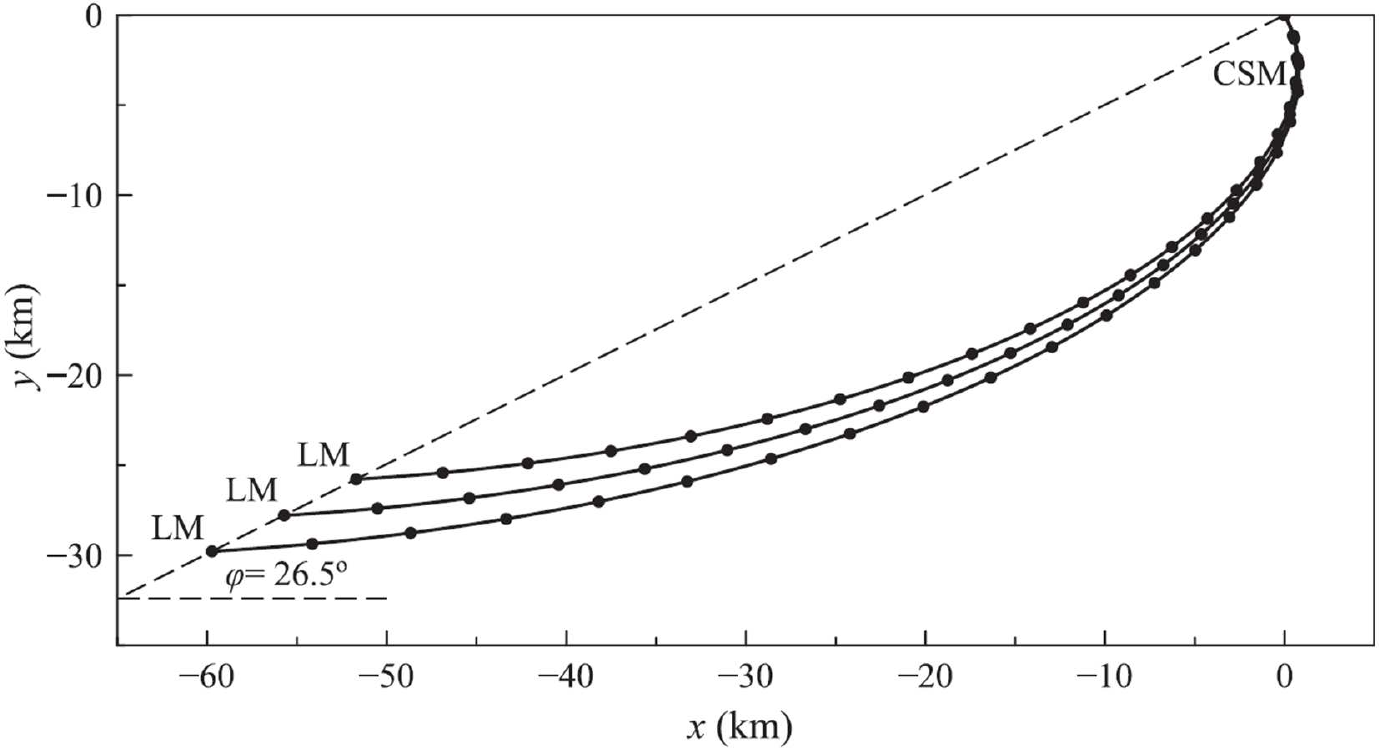}
  \caption{The rendezvous of the lunar module (LM) with the command/service module (CSM) of Apollo 11 from three initial values of $y_0$.  In each case, the LM sees the CSM at the same elevation angle $\phi$ at the same time.  The dots mark the LM's positions at 2-min intervals.}
\label{fig:phi}
\end{figure}

The timeline of the Apollo 11 flight plan was very tight.  The triggering of the TPI when the CSM appeared at an elevation angle of 26.5$^\circ$ above the LM's local horizon helped keep to that timeline.  The LM could have been in an orbit that was too high or too low.  Fig.~\ref{fig:phi} shows the command/service module and three possible heights for the LM's orbit.  Each LM sees the CSM at the same elevation angle $\phi$ above the positive $x$-axis so $x_0=y_0\cot(\phi)$.  If this is substituted for $x_0$ in Eqs.~(\ref{eq:xdot0}) and (\ref{eq:ydot0}), then $\dot{x}_0$, $\dot{y}_0$, and $v_{{\rm pre},x}$ are all proportional to $y_0$ (and $v_{{\rm pre},y}=0$).  Thus Eq.~(\ref{eq:thetaaim2}) shows that for a given flight time $t_f$ and elevation angle $\phi$, the aiming angle is independent of the height of the LM's orbit.  Although the values of $\Delta v_x$ and $\Delta v_y$ will vary, they are both proportional to $y_0$, which makes it simpler to make adjustments to the magnitude of the thrust.  Furthermore, Fig.~\ref{fig:phi} shows that at any later time, the three LMs always observe the CSM at the same elevation angle.  This simplified the astronauts' search for the other spacecraft because it would always appear at the same angular position at a given time.  The constancy of the time-of-flight and elevation angle $\phi$ helped the Apollo 11 mission adhere to its timeline, and meant that the LM would not have to change its aiming angle if it were higher or lower than its nominal orbit.  By taking advantage of the mathematical properties of these rendezvous equations, the Apollo mission planners were able to simplify the astronauts' tasks and make the most efficient use of their time.

\section{Summary}

As seen from the translating and rotating frame of a target spacecraft, the trajectory required for an interceptor spacecraft to coast to a rendezvous with it is never a straight line.  The linear path followed between orbiting spacecraft or astronauts in some movies and books is not realistic because non-inertial ``tidal'' and Coriolis forces cause the coasting interceptor to follow a drifting ellipse.  The sum of the ``tidal'' and Coriolis forces acting on the interceptor is a Hooke's law force that is directed toward the instantaneous center of the ellipse.  Even staying at rest relative to one another is not possible (without using thrusters) if the two spacecraft are not following the same orbit.  It is only when the target and interceptor are close (within 40~m or so) that reasonably accurate line-of-sight targeting is possible.  The lunar rendezvous of the Apollo 11 mission provides a good example of how to take advantage of the properties of the linearized equations of motion in the non-inertial frame of the target spacecraft.  The results reproduce well the actual rendezvous trajectory of the lunar module with the orbiting command/service module.  Now, fifty years after Apollo 11, eight nations have plans to return to the Moon, with the United States, China, and Japan contemplating landing men and women on the lunar surface.  Plans for additional missions to Mars and its two moons are also underway.  Some of these missions, especially the crewed missions, will involve the rendezvous of a lander with an orbiting spacecraft, and those missions will use the same techniques outlined here.

\begin{acknowledgments}

The author greatly appreciates the helpful comments of his two referees, Dr.\ Michael Efroimsky of the U.S.\ Naval Observatory and another unknown to him.  The author would also like to thank Dr.\ Jennifer Levasseur of the National Air and Space Museum for her help in obtaining information about Fig.~\ref{fig:rendezvous}.  The author is grateful for the continued support of the Physics Department at Weber State University.

Published by the American Journal of Physics {\bf 87}, 627 (2019); $\langle$\url{https://doi.org/10.1119/1.5115341}$\rangle$.  

\end{acknowledgments}


\begin{thebibliography}{99}

\bibitem{Hansen} James R.\ Hansen, ``The rendezvous that was almost missed: Lunar orbit rendezvous and the Apollo program,'' NASA Facts 175 (1992).

\bibitem{HackerGrimwood1} Barton C.\ Hacker and James M.\ Grimwood, {\em On the shoulders of titans}, NASA SP-2403 (NASA Science and Technical Information Office, 1977), p.\ 246.

\bibitem{HackerGrimwood2} Barton C.\ Hacker and James M.\ Grimwood, {\em On the Shoulders of Titans}, NASA SP-2403 (NASA Science and Technical Information Office, 1977), p.\ 288.

\bibitem{Wiesel} William E.\ Wiesel, {\em Spaceflight Dynamics}, 3rd ed.\ (Aphelion Press, Beavercreek, OH, 2010), p.\ 89.

\bibitem{Prussing} John E.\ Prussing and Bruce A.\ Conway, {\em Orbital Mechanics}, 2nd ed.\ (Oxford U.P., Oxford, UK, 2013), p.\ 27.

\bibitem{Freedman} Roger A.\ Freedman, Ibrahim Helmy, and Peter D.\ Zimmerman, ``Simplified navigation for self-propelled astronauts,'' Am.\ J.\ Phys.\ {\bf 43}, 438 -- 440 (1975).

\bibitem{Butikov} Eugene I.\ Butikov, ``Relative motion of orbiting bodies,'' Am.\ J.\ Phys.\ {\bf 69}, 63 -- 67 (2001).

\bibitem{Amato} Joseph C.\ Amato, ``Flying in formation: The orbital mechanics of LISA's three spacecraft,''  Am.\ J.\ Phys.\ {\bf 87}, 18 -- 23 (2019).

\bibitem{Mueller} Donald D.\ Mueller, ``Relative Motion in the Docking Phase of Orbital Rendezvous,'' Technical Documentary Report No. AMRL-TDR-62-124 (Behavioral Sciences Laboratory, Wright-Patterson Air Force Base, OH, 1962).

\bibitem{Mills} Blake D.\ Mills Jr., ``Satellite Paradox,'' Am.\ J.\ Phys.\ {\bf 27}, 115 -- 117 (1959). 

\bibitem{Omran} Ashraf Omran and Brett Newman, ``Nonlinear Analytical Solution of Relative Motion Subject to J2 Perturbation Using Volterra Kernels,'' Adv.\ Astronaut.\ Sci.\ {\bf 142}, 987 -- 1006 (2012).

\bibitem{Lovell} T.\ Alan Lovell, ``Analysis of a New Nonlinear Solution of Relative Orbital Motion (Abstract),'' in 23rd International Symposium on Space Flight Dynamics (2012), $\langle$\url{http://issfd.org/ISSFD_2012/ISSFD23_FF1_4_abstract.pdf}$\rangle$ (accessed May 1, 2019).
  
\bibitem{Stringer} Mary T.\ Stringer, Brett Newman, T.\ Alan Lovell, and Ashraf Omran ``Analysis of a New Nonlinear Solution of Relative Orbital Motion'' in 23rd  International Symposium on Space Flight Dynamics (2012), $\langle$\url{http://issfd.org/ISSFD_2012/ISSFD23_FF1_4.pdf}$\rangle$ (accessed May 1, 2019).

\bibitem{Symon} Keith  R.\ Symon, {\em Mechanics}, 2nd ed.\ (Addison-Wesley, Reading, MA, 1960), p.\ 278.

\bibitem{Carroll} Bradley W.\ Carroll and Dale A.\ Ostlie, {\em An Introduction to Modern Astrophysics}, 2nd ed.\ (Addison-Wesley, San Francisco, CA, 2007), p.\ 49.

\bibitem{Hill} G.\ W.\ Hill, ``Researches in the Lunar Theory,'' Am.\ J.\ Math.\ \textbf{1}(1), 5 -- 26 (1878).

\bibitem{CW} W.\ H.\ Clohessy and R.\ S.\ Wiltshire, ``Terminal Guidance System for Satellite Rendezvous,'' J.\ Aerosp. Sci.\ \textbf{27}(9), 653 -- 658 (1960).

\bibitem{Meeus} Jean Meeus, {\em Astronomical Algorithms}, 2nd ed.\ (Willmann-Bell, Inc., Richmond, VA, 1998), p.\ 237. 

\bibitem{Aldrin} Edwin E. Aldrin, ``Line-of-Sight Guidance Techniques for Manned Orbital Rendezvous,'' Ph.D.\ thesis (M.I.T., 1963), p.\ 45.

\bibitem{Collins} Michael Collins, {\em Carrying the Fire} (Farrar, Straus, and Giroux, New York, NY, 1974), p.\ 209.

\bibitem{Michalik} Thomas R.\ Michalik, ``Reference frames and orbital rendezvous,'' Phys.\ Teach.\ {\bf 23}, 253 -- 254 (1985).

\bibitem{Journal2} ``Apollo Flight Journal,'' $\langle$\url{https://www.history.nasa.gov/afj/ap11fj/19day6-rendezvs-dock.html}$\rangle$ (accessed May 1, 2019).

\bibitem{NASA} ``Project Apollo: 'Lunar Orbit Rendezvous' (1968) NASA Mission Planning and Analysis Division,'' 
$\langle$\url{https://www.youtube.com/watch?v=eo3D6gU9csw}$\rangle$ (accessed May 1, 2019).

\bibitem{Apollo11} ``Apollo 11 Flight Plan: Final July 1, 1969,''  $\langle$\url{https://www.hq.nasa.gov/alsj/a11/a11fltpln_final_reformat.pdf}$\rangle$ (accessed May 1, 2019).

\bibitem{Journal}``Apollo Flight Journal,'' $\langle$\url{http://www.history.nasa.gov/afj/ap10fj/as10-day5-pt21.html}$\rangle$ (accessed May 1, 2019). 

\bibitem{Apollovideo} ``Apollo 11 -- 16-mm magazine 1122-D,''  $\langle$\url{https://www.youtube.com/watch?v=TB4TenTk-Bc}$\rangle$ (accessed May 1, 2019).

\end{thebibliography}
\end{document}